\documentclass[aps,prl,superscriptaddress,twocolumn,citeautoscript]{revtex4-1}
\usepackage{graphicx}
\usepackage{amsfonts}
\usepackage{amsmath}
\usepackage{amssymb}
\usepackage{bm}
\usepackage{textcomp}
\usepackage{verbatim}

\usepackage{soul}

\usepackage{bbold}
\usepackage{mathtools}
\usepackage{dcolumn}
\usepackage{hyperref}
\usepackage[usenames]{color}
\usepackage{subfig}

\usepackage{physics}
\usepackage{units}
\usepackage[export]{adjustbox}

\usepackage{tablefootnote}

\usepackage{xcolor}

\usepackage{ragged2e}
\DeclareCaptionJustification{justified}{\justifying}
\def\beq{\begin{equation}}
\def\eeq{\end{equation}}

\def\vR{{\bf R}}

\def\vq{{\bf q}}

\newcommand{\out}[1]{{}}

\begin{document}

\title{Ferro-octupolar  order and low-energy excitations in d$^2$ double perovskites of Osmium}

\author{Leonid V. Pourovskii}
\address{Centre de Physique Th\'eorique, Ecole Polytechnique, CNRS, Institut Polytechnique de Paris, 91128 Palaiseau Cedex, France}
\address{Coll\`ege de France, 11 place Marcelin Berthelot, 75005 Paris, France}

\author{Dario Fiore Mosca}
\address{University of Vienna, Faculty of Physics and Center for Computational Materials Science, 1090 Vienna, Austria}

\author{Cesare Franchini}
\address{University of Vienna, Faculty of Physics and Center for Computational Materials Science, 1090 Vienna, Austria}
\address{Department of Physics and Astronomy "Augusto Righi", Alma Mater Studiorum - Universit\`a di Bologna, Bologna, 40127 Italy}

\begin{abstract}
Conflicting interpretations of experimental data preclude the understanding of the quantum magnetic state of 
spin-orbit coupled $d^2$ double perovskites. Whether the ground state is a Janh-Teller-distorted order of quadrupoles 
or the hitherto elusive octupolar order remains debated. We resolve this uncertainty through direct calculations of all-rank inter-site exchange interactions and inelastic neutron scattering  cross-section 
for the $d^2$ double perovskite series 
Ba$_2M$OsO$_6$  ($M$= Ca, Mg, Zn). Using advanced many-body first principles methods we show that the ground state is formed by ferro-ordered octupoles coupled {
by superexchange interactions} within the ground-state $E_g$ doublet. Computed ordering temperature of the single second-order phase-transition 
{ is consistent with experimentally observed material-dependent trends.}
Minuscule distortions of the parent cubic structure are shown to qualitatively modify the structure of gaped magnetic excitations. 
\end{abstract}


\maketitle

Identification of complex magnetic orders in spin-orbital entangled and electronically correlated transition metal oxides has emerged as a fascinating field of study, enabling the discovery of new quantum magnetic states originating from 
interaction between effective pseudospins 
carrying high-rank multipoles~\cite{doi:10.7566/JPSJ.90.062001,Witczak-Krempa2014}. 
{ 
While multipolar coupling in localized f-electrons systems has been the subject of intense research and is overall well understood~\cite{10.1143/PTPS.176.77,Santini2009}, the formation and quantum origin of ordered multipoles in $d$-electron systems is a much more recent research area which poses challenging issues and controversial opinions.~\cite{PhysRevLett.115.026401, PhysRevLett.103.067205, PhysRevLett.115.026401, Maharaj2020,doi:10.1126/science.aad1188,khaliullin2021}. 
Ordered magnetic octupoles were initially proposed as an alternative orbital ordering in $e_g$ manganites arising from the complex mixing of doubly-degenerate orbitals~\cite{doi:10.1143/JPSJ.69.3328, PhysRevB.63.140416}, and later in spin-orbit coupled model systems 
analogous to Sr$_2$VO$_4$, LiOsO$_3$ and Cd$_2$Re$_2$O$_7$ 
~\cite{PhysRevLett.103.067205,PhysRevLett.115.026401, doi:10.1126/science.aad1188}. 
Rock-salt ordered double perovskites (DP) Ba$_2M$OsO$_6$ ($M$=Ca, Mg, Zn; in short: BCOO, BMOO and BZOO) represent the first candidate materials experimentally proposed to host a  $d$-orbital octupolar order~\cite{Maharaj2020}. However, the possibility to actually realize such an exotic magnetic order and the driving mechanism responsible for its formation remain largely debated, 
in particular regarding the rank of the multipolar interactions at play, the degree of JT distortions and the relative importance of  direct and indirect exchange~\cite{Maharaj2020, Paramekanti2020, PhysRevB.101.155118, PhysRevB.102.064407, doi:10.7566/JPSJ.90.062001, khaliullin2021}.
}

In these Os-based DPs the strong spin-orbit coupling (SOC) strength splits the effective $L=1$ $t_{2g}$ levels on the magnetic Os ions into a lower $j=\frac{3}{2}$ quadruplet ground state (GS) and a doublet $j=\frac{1}{2}$ excited state. With a $d^2$ (S=1) configuration, the total angular momentum $J_{eff}$ is 2, and the levels are split due to the remnant crystal field (CF) into a lower $E_g$ doublet and $T_{2g}$ triplet~\cite{PhysRevB.84.094420,Maharaj2020}. In contrast to the assumptions of the pioneering theoretical study of Ref.~\onlinecite{PhysRevB.84.094420}, the intersite exchange interactions 
are inferred to be much smaller than the remnant CF~\cite{Yamamura2006,Maharaj2020}. 
Despite experimental evidence for a single phase transition involving the $E_g$ manifold~\cite{Maharaj2020, Thompson2014, Marjerrison2016}, its origin remains unclear.

Considering that the non-Kramers $E_g$ doublet does not carry dipole moments it would be legitimate to expect that  conventional quadrupolar couplings in a JT-broken symmetry would promote an anti-ferro (AF) quadrupolar  order~\cite{khaliullin2021, Churchill2021}. This transparent picture does not seem to be consistent with recent experiments: X-ray diffraction (XRD) does not find structural distortions (larger than 0.1$\%$) and, whereas no conventional magnetic order is detected by neutron diffraction (upper limit $\approx$ 0.1 $\mu_B$), muon spin relaxation still indicates time-reversal (TR) symmetry breaking thereby ruling out quadrupolar order~\cite{Marjerrison2016}. To account for the experimental measurements a ferro-octupolar (FO) ordered GS was proposed~\cite{Maharaj2020, Paramekanti2020, PhysRevB.101.155118, Voleti2021}, involving a coupling between the lower $E_g$ and excited $T_{2g}$ states mediated by quadrupolar operators. 
This 
model reproduces a spin-gap observed in the excitation spectra~\cite{Maharaj2020,Paramekanti2020} and is overall reasonably compatible with the experimental scenario, but it makes use of some problematic assumptions. Only a subset of inter-site exchange interactions (IEI) allowed within $J_{eff}$=2 is assumed to be non-zero. Moreover, the included quadrupole IEI, which cannot be directly  inferred from experiment, are tuned to obtain the desired properties of the FO phase. 

Inspired by the apparent adequacy of the experimentally proposed FO state 
{ and aiming to decipher the key aspects of FO ordering in 5$d$-electron systems}
we propose in this Letter an alternative mechanism based on a direct numerical calculations of all possible interaction channels by means of many-body first principles schemes.
Without forcing any pre-assumption on the form of the effective Hamiltonian we find a ferro order of $xyz$ octupoles determined by a competition between time-even and octupolar IEI within solely the GS $E_g$ doublet. 
{ Importantly, employing an analysis that discriminates between direct exchange (DE) and superexchange (SE) mechanisms we found that IEI are dominated by SE through O-2$p$ and Ba orbitals; 5$d$-5$d$ DE contributes only marginally.}  
Our data correctly predict the observed  second-order phase transition, with computed ordering temperature compatible with the experimental one, and a gapped excitation spectra.

{\it Effective Hamiltonian and methods.}  
The effective Hamiltonian for the low-energy degrees of freedom on the Os sublattice is a sum of the IEI ($H_{IEI}$) and remnant crystal-field (rcf) terms:
\beq\label{eq:Heff}
H_{eff}=\sum_{\langle
ij\rangle}\sum_{KQK'Q'}V_{KK'}^{QQ'}(\Delta\vR_{ij})O_K^Q(\vR_i)O_{K'}^{Q'}(\vR_j) +\sum_i H^i_{rcf},
\eeq
where the first sum 
is over all $\langle ij\rangle$ Os-Os bonds, $O_K^Q(\vR_i)$ is the Hermitian spherical tensor~\cite{Santini2009} for $J$=2  of the rank $K=1...4$ and projection $Q$ acting on Os site at the position $\vR_i$,   the  IEI $V_{KK'}^{QQ'}(\Delta\vR_{ij})$ acts between the multipoles $KQ$ and $K'Q'$ on two Os sites connected by the lattice vector $\Delta\vR_{ij}=\vR_j-\vR_i$. Finally, 
$H^i_{rcf}=-V_{rcf}\left[\mathcal{O}_4^0(\vR_i)+5 \mathcal{O}_4^4(\vR_i)\right]$ is the 
remnant octahedral CF \cite{Maharaj2020}, where  $\mathcal{O}_K^Q$ are the standard Stevens operators. 

To derive the above Hamiltonian we use density functional theory (DFT)~\cite{Wien2k} + dynamical mean-field theory (DMFT)~\cite{Georges1996,Anisimov1997_1,Lichtenstein_LDApp,Aichhorn2016}  { treating the quantum impurity problem on the Os 5$d$ shells} within the quasi-atomic Hubbard-I (HI) approximation~\cite{hubbard_1}. 
All IEI $V_{KK'}^{QQ'}(\Delta\vR)$ 
are computed 
within the HI-based force-theorem approach (FT-HI)~\cite{Pourovskii2016}.
Our DFT+HI calculations correctly predict the expected $J_{eff}=$ 2 GS multiplet, which is split by $H_{rcf}$ into the ground state $E_g$ doublet and excited $T_{2g}$ triplet. 
More details can be found in the Supplementary Materials (SM)~\cite{supplmat}.

\begin{figure}[!tb]
  	\begin{centering}
  		\includegraphics[width=0.9\columnwidth]{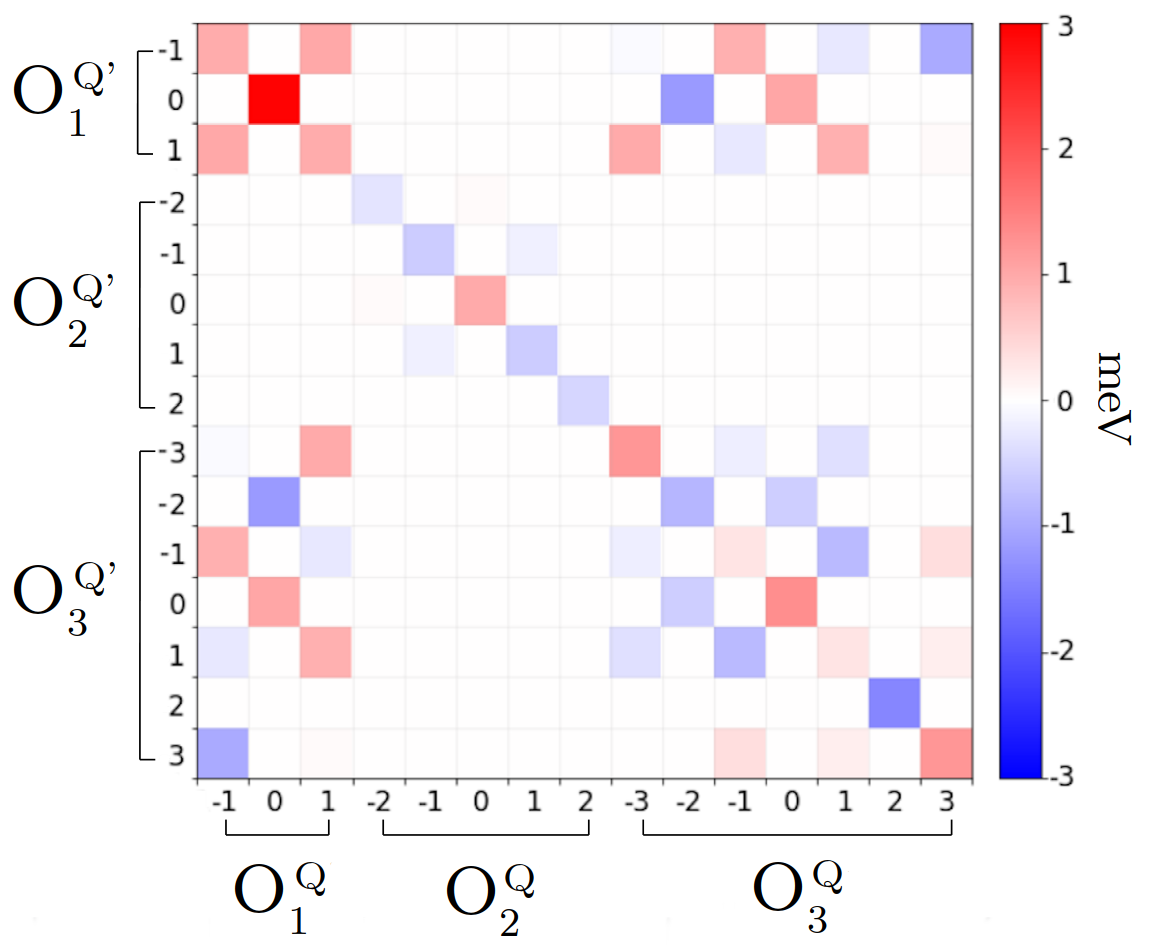} 
  		\par\end{centering}
  	\caption{Color map of the inter-site exchange interactions (IEI) $V_{KK'}^{QQ'}$, eq.~~\ref{eq:Heff},  in BZOO for the [1/2,1/2,0] Os-Os pair. The IEI involving hexadecapoles ($K$=4) are negligible and not included. The complete list of $V_{KK'}^{QQ'}$ for the three compounds is given in the SM~\cite{supplmat}.}
  	\label{fig:matrix} 
  \end{figure}

{\it CF splitting and intersite exchange
interactions.}  The  calculated  CF splitting $\Delta_{rcf}=120 V_{rcf}$ 
listed in Table~\ref{Tab:cf_SEI} is about 20 meV for all members, in agreement with specific heat measurements and  excitation gap inelastic neutron scattering (INS) 
~\cite{Yamamura2006,Maharaj2020,Paramekanti2020}. 
The computed IEI $V_{KK'}^{QQ'}$ are displayed in Fig.~\ref{fig:matrix} (for BZOO, similar data are obtained for the other members, see SM). The largest values, $\approx 3$~meV are significantly smaller than $\Delta_{rcf}$, in agreement with experiment~\cite{Maharaj2020,Paramekanti2020}, implying that the ordered phase  will be determined by the IEI acting within the ground-state $E_g$ doublet.

\begin{table}[!b]
	\begin{center}
		\begin{ruledtabular}
			\renewcommand{\arraystretch}{1.2}
			\begin{tabular}{l c c c c }
				Compound & $\Delta_{rcf}$  & $J_{yy}$ &  $J_{zz}$  & $J_{xx}$  \\
				\hline
				Ba$_2$CaOsO$_6$ & 17.1 & -2.98 & 1.48 & -0.61 \\
				
				Ba$_2$MgOsO$_6$ & 19.2 & -2.93 &  1.67 & -0.69  \\
				
				Ba$_2$ZnOsO$_6$ & 20.5 & -1.71& 1.35  & -0.50 \\
				
			\end{tabular}
		\end{ruledtabular}
		\caption{\label{Tab:cf_SEI}  Remnant CF splitting $\Delta_{rcf}$ and IEI  $J_{\alpha\alpha}$  
		within the $E_g$ doublet for the Os-Os [1/2,1/2,0] lattice vector. All values are in meV.}
	\end{center}
\end{table}

The $E_g$ space can be encoded by spin-1/2 operators $\tau_{\alpha}$, { with the $E_g$ states corresponding to the projections $\pm1/2$ of pseudo-spin-1/2.
The resulting $E_g$ Hamiltonian 
\beq\label{eq:H_Eg}
H_{E_g}=\sum_{\langle ij\rangle \in NN}\sum_{\alpha\beta}J_{\alpha\beta}(\Delta \vR_{ij})\tau_{\alpha}(\vR_i)\tau_{\beta}(\vR_j),
\eeq
 is %
 eq.~\ref{eq:Heff} downfolded into the $E_g$ space  (see SM~\cite{supplmat} for details).}
Up to a normalization factor, $\tau_y$ is the octupole  $O_3^{-2}\equiv  O_{xyz}$; the corresponding IEI $V_{33}^{\bar{2}\bar{2}}$ directly maps into $J_{yy}$.  $\tau_x$ and $\tau_z$ are combinations of the $e_g$ quadrupoles ($O_2^2$ and $O_2^0$, respectively) with  hexadecapoles of the same symmetry. Therefore, $V_{22}^{22}$ and $V_{22}^{00}$   together with the corresponding hexadecapole IEI contribute to $J_{xx}$ and $J_{zz}$, respectively. Since the hexadecapole IEI are negligible 
their admixture reduces the magnitude of time-even $J_{xx}$ and $J_{zz}$  (Sec.~III in SM~\cite{supplmat}).
 Overall, the order in $E_g$ space is determined by a competition of the time-even ($\tau_x$ and $\tau_z$) combinations of quadrupoles and hexadecapoles with the time-odd $xyz$ octupole. There are, correspondingly, no IEI coupling $\tau_y$ with $\tau_x$ or $\tau_z$ due to their different symmetry under the time reversal.

\begin{figure}[!tb]
  	\begin{centering}
  		\includegraphics[width=0.95\columnwidth]{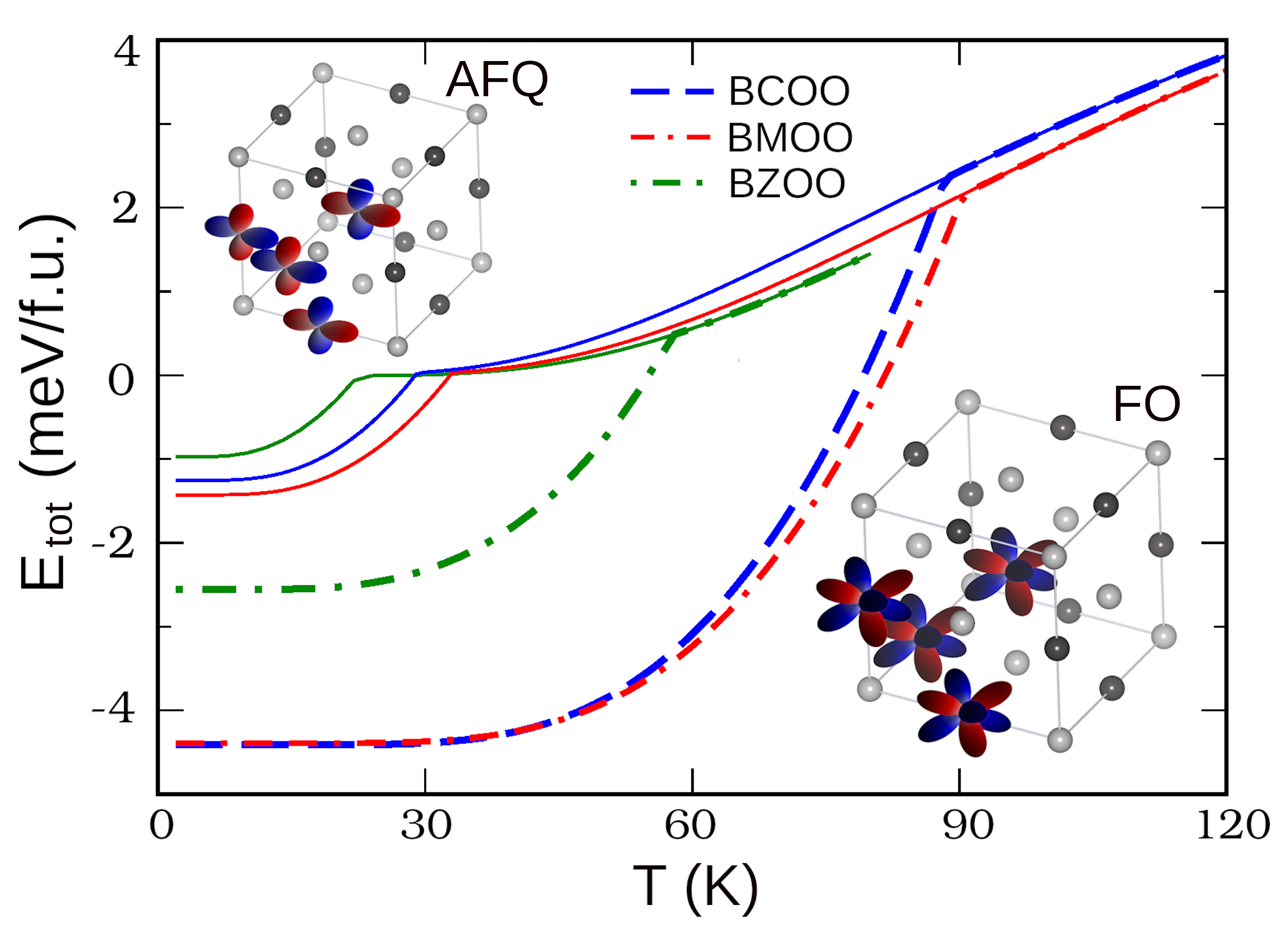} 
  		\par\end{centering}
  	\caption{Mean-field 
  	total energy vs. temperature calculated from the Hamiltonian (\ref{eq:Heff}), with the zero energy corresponding to the ground state energy of $H_{rcf}$ ($E_g$ doublet). The bold lines are the energies calculated from the full Hamiltonian. The thin solid lines of the corresponding colors are calculated with the IEI between $xyz$ octupoles set to zero.
  	 { The insets depict the resulting FO and AFQ orders.}}
  	\label{fig:E_MF} 
  \end{figure}

  Our calculated $E_g$ IEI for the [1/2,1/2,0] lattice vector are listed in Table~\ref{Tab:cf_SEI}. There are no off-diagonal couplings in this case -- only $J_{\alpha\alpha}$ are non-zero. The IEI for other NN lattice vectors are obtained by transforming ($\tau_x$,$\tau_z$) with the corresponding rotation matrices of the $e_g$ irreducible representation; $J_{yy}$ is the dominant interaction and, as expected, the same for all the NN bonds; its negative sign corresponds to a ferromagnetic coupling between $xyz$ octupoles, as schematically shown in the inset of Fig.~\ref{fig:E_MF}. The magnitude of $J_{yy}$ varies substantially between the systems, being about 40\% smaller in BZOO as compared to BMOO or BCOO.
  The IEI in the time-even ($\tau_x$,$\tau_y$) space are smaller and positive (AF), leading to a possible frustration on the fcc Os sublattice.
  
  We note that our results are qualitatively different from previous assumptions  \cite{Paramekanti2020,khaliullin2021}, since  we obtain a significant value for the $xyz$ octupolar IEI  $V_{33}^{\bar{2}\bar{2}}$ in the $J_{eff}$ space, see Fig.~\ref{fig:matrix}. Since the $xyz$ octupole is directly mapped to $\tau_y$, this results in  large $J_{yy}$. In contrast, Ref.~\onlinecite{Paramekanti2020} assumed zero $V_{33}^{\bar{2}\bar{2}}$; to obtain a resonable value for effective $J_{yy}$  through an "excitonic" mechanism,  a huge quadrupole IEI $V_{xy-xy}\equiv V_{22}^{\bar{2}\bar{2}}\sim$35~meV (in our spherical tensor normalization) was employed, which is about 2 orders of magnitude larger than the one predicted by our calculations (see Fig.~\ref{fig:matrix} and SM~ ~\cite{supplmat}). Ref.~\onlinecite{khaliullin2021} considered only Os-Os DE and found the  $J_{yy}$ IEI to be zero. 
  
  {In order to discriminate between  DE and various SE contributions to the IEI we have developed an approach to exclude a chosen set of hopping processes from IEI. This approach is based on expanding the downfolded Os 5$d$ orbitals onto a 
  set of all relevant valence states (Ref.~\cite{Delange2017}, see SM \cite{supplmat} for details). 
  This analysis shows that the effect of Os-Os DE is insignificant (below 10\%). The IEI are dominated by SE processes, involving hoppings through O-2p and Ba states (Supp. Table II~\cite{supplmat} ), with contributions of similar magnitude to both quadrupolar and octupolar IEI. These results explain the comparable strength of quadrupolar and octupolar IEI in the $J_{eff}=2$ space (Fig.~\ref{fig:matrix} and SM~\cite{supplmat}). The time-even IEI in the $E_g$ space are then further diminished  by the hexadecapoles admixture into $\tau_x$ and $\tau_z$ as discussed above, resulting in a dominating $xyz$ coupling $J_{yy}$ (Table~\ref{Tab:cf_SEI}).}
  
  { A dominating SE  also  naturally explains the weaker IEI in BZOO as compared to two other systems. Substituting Mg or Ca at the $M$ site by more electronegative Zn results in a  more covalent $M$-O bond that weakens the Os-O bond through  "covalency competition"\cite{Yamada2017}. In result, the principal Os-Os SE coupling through O and Ba is reduced.}
  
  {\it Ordered phase.}  
  From the first-principles effective Hamiltonian (\ref{eq:Heff}) we evaluate  the ordered phases and transition temperatures $T_o$  within the mean-field approximation (MFA) 
  ~\cite{Rotter2004}. All three systems exhibit a single 2$^\mathrm{nd}$ order  phase transition into the FO $xyz$ phase, as shown in Fig.~\ref{fig:E_MF} where the zero-T limit corresponds to the FO  ground-state ordering energy. The only non-zero $J_{eff}=$2 multipoles  at the FO ground state are
  $\langle O_{xyz} \rangle$ (fully saturated at $1/\sqrt{2}$ for the spherical tensor normalization) as well as the "40" and "44" hexadecapoles arising due to  $H_{rcf}$ and 
  exhibiting no peculiarity at $T_o$.  
  The quasi-linear behavior of $E_{tot}$ above the $T_o$  is due to the CF term. 
  The calculated values of the FO $T_o$ ($T_o^{\mathrm{FO}}$ in Table~\ref{Tab:To_ext}) systematically overestimate the experimental one by about 80\% due to the employed approximations (MFA and HI), in line with previous applications of the FT-HI framework~\cite{Horvat2017,Pourovskii2019,Pourovskii2021}, but the material dependent changes are captured very well
  (${T_o^{\mathrm{FO}_{BCOO}}}/{T_o^{\mathrm{FO}_{BZOO}}} \approx 1.6$, while ${T_o^{\mathrm{FO}_{BCOO}}}/{T_o^{\mathrm{FO}_{BMOO}}} \approx 1$).

  To explore competing time-even orders, we set the $xyz$ IEI to zero and obtain a planar AF order of the $e_g$ quadrupoles and associated hexadecapoles, with ferro-alignment of all  order parameters (encoded by $\langle \tau_x\rangle$ and $\langle \tau_z\rangle$)  within (001) planes that are AF-stacked in the [001] direction. {This structure (shown as inset in Fig.~\ref{fig:E_MF} as well as in SM~\cite{supplmat}) is in agreement with the quadrupolar order previously predicted by Ref.~\cite{khaliullin2021}}.  The corresponding ordering temperature $T_o^{\mathrm{AF}}$ are about 3 times smaller than $T_o^{\mathrm{FO}}$ (see Fig.~\ref{fig:E_MF} and Table~\ref{Tab:To_ext})
  Considering that this AF order in the cubic phase is unstable against JT distortions~\cite{khaliullin2021}, the release of JT modes is expected to further stabilise the AF phase, but most unlikely by a factor of 3. 
  No sign of JT distortions above 0.1\% have been measured in BCOO~\cite{Maharaj2020}.

  \vspace{0.3cm}
  {\it Generalized susceptibility and on-site excitations.} 
Information on the characteristic excitations  of the FO $xyz$ order is encoded in generalized dynamical lattice (${\chi}(\vq,E)$) and single-site ($\chi_{0}(E)$) susceptibility, that we computed within the random phase approximation (RPA), see
Ref.~\onlinecite{RareEarthMag_book} and SM~\cite{supplmat}. The matrix elements  $\chi_{0}^{\mu\mu'}(E)$ are evaluated from the eigenvalues $E$ and eigenstates $\Psi$ of the $J_{eff}$=2 manifold:

  \beq\label{eq:chi_loc}
  \chi_{0}^{\mu\mu'}(E)=\sum_{AB}\frac{\langle \Psi_A|O_{\mu}|\Psi_B\rangle \langle \Psi_B|O_{\mu'}|\Psi_A\rangle}{E_B-E_A-E}\left[p_A-p_B\right],
  \eeq
  where  $A(B)$ labels five single-site eigenvalues and eigenstates of the  Hamiltonian  (\ref{eq:Heff}), the combined index $\mu=[K,Q]$  labels $J_{eff}$ multipoles, and  $p_{A(B)}$ is the corresponding Boltzmann weight. 
  
  \begin{table}[!t]
  	\begin{center}
  		\begin{ruledtabular}
  			\renewcommand{\arraystretch}{1.2}
  			\begin{tabular}{l c c c c c }
  				Compound & $T_o^{\mathrm{FO}}$ & $T_o^{\mathrm{AF}}$ & $T_o^{\mathrm{exp}}$ &  $E_S$  & $E_T$  \\
  				\hline
  				Ba$_2$CaOsO$_6$  & 89 & 29 & 49 & 17.7 & 25.9 \\
  				
  				Ba$_2$MgOsO$_6$ &  91 & 33 &  51 & 17.6 & 28.0  \\
  				
  				Ba$_2$ZnOsO$_6$ & 58 & 23 &  30 & 10.2 & 25.6 \\
  				
  			\end{tabular}
  		\end{ruledtabular}
  		\caption{\label{Tab:To_ext}  Calculated mean-field ordering temperatures $T_o$ (in K) for the FO $xyz$ and  time-even antiferro (AF) phases compared to the experimental values from Refs.~\onlinecite{Thompson2014,Marjerrison2016}. Last two columns: the energies (in meV) of the singlet ($E_S$) and triplet ($E_T$) excited levels of the $J_{eff}$=2 multiplet in the FO $xyz$ ground state.
  	    }
  	\end{center}
  \end{table}

  In the FO GS the $J_{eff}$=2 manifold is split into 3 levels (Table~\ref{Tab:To_ext}): singlet GS, first singlet (S) excited state (with opposite sign of $xyz$ octupole compared to GS and energy proportional to IEI) and a high-energy $T_{2g}$ triplet (T) due to $\Delta_{rcf}$ further enhanced by IEI (cf.  Tab.~\ref{Tab:cf_SEI}). { In contrast to the $E_g$ doublet, the $T_{2g}$ triplet degeneracy is not lifted by the $xyz$ exchange field, since the direct product $T_{2g}\times T_{2g}$ does not contain the irreducible representation $A_{2u}$ of the $xyz$ octupole.}
  
\begin{figure*}[!bth]
	\begin{centering}
		\includegraphics[width=2.0\columnwidth]{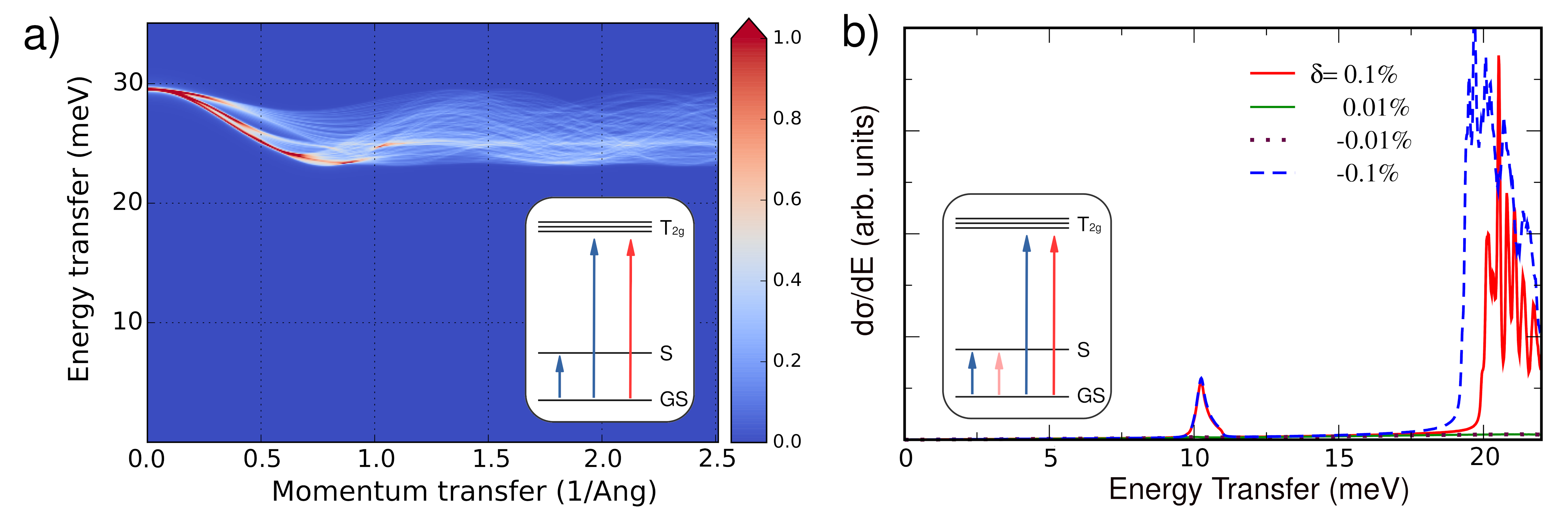} 
		\par\end{centering}
	\caption{(a) Color map of the calculated powder-averaged INS differential cross-section in cubic BZOO as a function of the energy transfer $E$ and momentum transfer $q$. Inset: on-site splitting of $J_{eff}$=2 levels in the ferro-octupolar phase. Allowed time-even and time-odd transitions between the levels are schematically shown by blue and red arrows, respectively. (b)  $q$-integrated INS differential cross-section of BZOO for the tetragonal distortions $\delta=\pm$0.1\% and $\pm$0.01\%. An exchange peak at about 10~meV is  clearly seen for the larger distortion. The onset of crystal-field excitations is seen above 18-20 meV. Inset shows the corresponding $J_{eff}$=2 level scheme with a time-odd ($xyz$) transition (pale red arrow) between the ground-state (GS) and singlet (S) levels turned on by the distortions.}
	\label{fig:INS_BZO} 
\end{figure*}

  
  We find that
  only $e_g$ quadrupoles and hexadecapoles connect the GS with the first excited S state, and since the IEI matrices do not couple time-odd and time-even multipoles, this S excitation can induce only time-even contributions to the RPA lattice susceptibility ${\chi}(\vq,E)$. In contrast, the matrix elements $\langle \Psi_{GS}|O_{\mu}|\Psi_{T}\rangle$ between GS and T levels take non-zero values for many odd and even multipoles (see inset in  Fig.~\ref{fig:INS_BZO}a).

{\it Inelastic neutron-scattering (INS) cross-section.} 
To provide further evidence directly comparable with available measurements~\cite{Maharaj2020}, from the knowledge of ${\chi}(\vq,E)$ we compute the magnetic contribution to the INS differential cross-section:
\begin{multline}\label{eq:INS_Xsec}
\frac{d^2 \sigma}{d \Omega dE'}\propto \sum_{\alpha\beta}\left(\delta_{\alpha\beta}-q_{\alpha}q_{\beta}\right) \\ \left[\sum_{\mu\mu'}F_{\alpha\mu}(\vq)F_{\beta\mu'}(\vq) \mathrm{Im} \chi_{\mu\mu'}(\vq,E)\right],
\end{multline}
where we drop unimportant prefactors. 
In order to take into account the octupole contributions into the INS cross-sections,  the form-factors $F_{\alpha\mu}(\vq)$ are evaluated beyond the dipole approximation on the basis of Refs.~\onlinecite{Shiina2007,Lovesey_book_full} (for more details see SM~\cite{supplmat}).

The calculated powder-averaged (averaged over $\vq$ directions) INS cross-section 
for BZOO is displayed in Fig~\ref{fig:INS_BZO}a (the similar results for BCOO and BMOO are given in SM). 
One clearly observes a band of CF excitations above 20~meV, in agreement with the magnitude of $E_T$. However, below the CF band one sees no features corresponding to transitions to the lower-energy S excitation.
As only odd-time multipoles contribute to the magnetic neutron scattering, this result can be anticipated from the structure of on-site excitations in the FO $xyz$ phase.

We conclude by showing the effect of minuscule tetragonal distortions $\delta$ on the INS spectrum. 
The remnant CF potential acting on the $J_{eff}$=2 multiplet  in the distorted structure becomes $H^i_{rcf}=-V_{rcf}\left[\mathcal{O}_4^0(\vR_i)+5 \mathcal{O}_4^4(\vR_i)\right]+V_t \mathcal{O}_2^0(\vR_i)$, where the tetragonal contribution $V_t=K_t \delta$.  
Using BZOO as case material, we perform a series of DFT+HI calculations for tetragonally-distorted BZOO for $\delta$ in the range -0.5 to 0.5\%   extracting $K_t=$266~meV (see SM).  Then, we add $\sum_i K_t \delta \mathcal{O}_2^0(\vR_i)$ to the Hamiltonian (\ref{eq:Heff}) and solve it in the MFA for small values of $\delta$ up to 0.1\%. 
We observe the same transition into the FO $xyz$ order with $T_o$ about 58~K as in the initial case. The only difference is that $\langle O_{z^2}\rangle $ is non-zero, reaching about 1/4 of its saturated value for $\delta=$0.1\% and an order of magnitude less for $\delta=$0.01\%. In the case of tetragonal compression ($\delta<$0) we obtain the same $\langle O_{z^2}\rangle $ magnitudes with opposite sign. The important point is that the GS and excited singlet $\Psi_S$ now feature non-zero matrix element for the time-odd $xyz$, $\langle \Psi_{GS}|O_{xyz}|\Psi_{S}\rangle  \propto  \langle O_{z^2}\rangle_{GS}$. Therefore, magnetic excitations across the gap become possible (see inset in Fig.~\ref{fig:INS_BZO}b) and should be, in principle, visible by INS. 

We evaluated the powder-averaged INS cross-section for a set of small distortions ($\delta=\pm$0.1\% and $\delta=\pm$0.01\%). We then integrate  $\frac{\delta^2 \sigma(q,\omega)}{d\Omega dE'}$ over the same range of $q$ and $E$ as the experimental INS spectra (Fig.~1 in Ref.~\onlinecite{Maharaj2020}). In the resulting cross-section shown in Fig~\ref{fig:INS_BZO}b the contribution of magnetic scattering across the exchange gap is completely negligible for $\delta=\pm$0.01\%. For the larger distortion ($\delta=\pm$0.1\%) a narrow peak emerges at $E\approx$10~meV, also visible in experimental INS data~\cite{Maharaj2020}. This peak has a small, but not negligible intensity as compared to the crystal-field excitations. The onset of the latter is shifting below 20~meV with increasing distortions (Fig.~\ref{fig:INS_BZO}b). 

  {\it Conclusions.} 
  Our first principles calculations provide robust qualitative and quantitative evidence  of a purely ferro order of $xyz$ octupoles in $d^2$ DPs~\cite{Maharaj2020,Paramekanti2020,Lovesey2020}, determined by 
  a competition between the time-even and octupolar IEI within the ground-state $E_g$ doublet, alternative to previous models based on unrealistically large quadrupolar coupling. { Our study reveals the role of superexchange as the main mechanism for triggering the formation of octupolar ordering in spin-orbit coupled 5$d$ oxides. 
  }
  The obtained ordering temperatures are consistent { with material-dependent trends}.  The simulated INS spectrum correctly reproduces the CF excitations in the cubic phase, and small tetragonal distortions are necessary to activate  the $O_{xyz}$ octupole operator 
  connecting the exchange-split ground and first excited states to generate the measured exchange peak~\cite{Maharaj2020}.

\begin{acknowledgements}
Support by the European Research Council grant ERC-319286-"QMAC" is gratefully  acknowledged. D. Fiore Mosca acknowledges the Institut Français d'Autriche and the French Ministry for Europe and Foreign Affairs for the French Government Scholarship. 
\end{acknowledgements}


\begin{thebibliography}{51}%
	\makeatletter
	\providecommand \@ifxundefined [1]{%
		\@ifx{#1\undefined}
	}%
	\providecommand \@ifnum [1]{%
		\ifnum #1\expandafter \@firstoftwo
		\else \expandafter \@secondoftwo
		\fi
	}%
	\providecommand \@ifx [1]{%
		\ifx #1\expandafter \@firstoftwo
		\else \expandafter \@secondoftwo
		\fi
	}%
	\providecommand \natexlab [1]{#1}%
	\providecommand \enquote  [1]{``#1''}%
	\providecommand \bibnamefont  [1]{#1}%
	\providecommand \bibfnamefont [1]{#1}%
	\providecommand \citenamefont [1]{#1}%
	\providecommand \href@noop [0]{\@secondoftwo}%
	\providecommand \href [0]{\begingroup \@sanitize@url \@href}%
	\providecommand \@href[1]{\@@startlink{#1}\@@href}%
	\providecommand \@@href[1]{\endgroup#1\@@endlink}%
	\providecommand \@sanitize@url [0]{\catcode `\\12\catcode `\$12\catcode
		`\&12\catcode `\#12\catcode `\^12\catcode `\_12\catcode `\%12\relax}%
	\providecommand \@@startlink[1]{}%
	\providecommand \@@endlink[0]{}%
	\providecommand \url  [0]{\begingroup\@sanitize@url \@url }%
	\providecommand \@url [1]{\endgroup\@href {#1}{\urlprefix }}%
	\providecommand \urlprefix  [0]{URL }%
	\providecommand \Eprint [0]{\href }%
	\providecommand \doibase [0]{http://dx.doi.org/}%
	\providecommand \selectlanguage [0]{\@gobble}%
	\providecommand \bibinfo  [0]{\@secondoftwo}%
	\providecommand \bibfield  [0]{\@secondoftwo}%
	\providecommand \translation [1]{[#1]}%
	\providecommand \BibitemOpen [0]{}%
	\providecommand \bibitemStop [0]{}%
	\providecommand \bibitemNoStop [0]{.\EOS\space}%
	\providecommand \EOS [0]{\spacefactor3000\relax}%
	\providecommand \BibitemShut  [1]{\csname bibitem#1\endcsname}%
	\let\auto@bib@innerbib\@empty
	\bibitem [{\citenamefont {Takayama}\ \emph {et~al.}(2021)\citenamefont
		{Takayama}, \citenamefont {Chaloupka}, \citenamefont {Smerald}, \citenamefont
		{Khaliullin},\ and\ \citenamefont {Takagi}}]{doi:10.7566/JPSJ.90.062001}%
	\BibitemOpen
	\bibfield  {author} {\bibinfo {author} {\bibfnamefont {T.}~\bibnamefont
			{Takayama}}, \bibinfo {author} {\bibfnamefont {J.}~\bibnamefont {Chaloupka}},
		\bibinfo {author} {\bibfnamefont {A.}~\bibnamefont {Smerald}}, \bibinfo
		{author} {\bibfnamefont {G.}~\bibnamefont {Khaliullin}}, \ and\ \bibinfo
		{author} {\bibfnamefont {H.}~\bibnamefont {Takagi}},\ }\href {\doibase
		10.7566/JPSJ.90.062001} {\bibfield  {journal} {\bibinfo  {journal} {Journal
				of the Physical Society of Japan}\ }\textbf {\bibinfo {volume} {90}},\
		\bibinfo {pages} {062001} (\bibinfo {year} {2021})}\BibitemShut {NoStop}%
	\bibitem [{\citenamefont {Witczak-Krempa}\ \emph {et~al.}(2014)\citenamefont
		{Witczak-Krempa}, \citenamefont {Chen}, \citenamefont {Kim},\ and\
		\citenamefont {Balents}}]{Witczak-Krempa2014}%
	\BibitemOpen
	\bibfield  {author} {\bibinfo {author} {\bibfnamefont {W.}~\bibnamefont
			{Witczak-Krempa}}, \bibinfo {author} {\bibfnamefont {G.}~\bibnamefont
			{Chen}}, \bibinfo {author} {\bibfnamefont {Y.~B.}\ \bibnamefont {Kim}}, \
		and\ \bibinfo {author} {\bibfnamefont {L.}~\bibnamefont {Balents}},\ }\href
	{\doibase 10.1146/annurev-conmatphys-020911-125138} {\bibfield  {journal}
		{\bibinfo  {journal} {Annual Review of Condensed Matter Physics}\ }\textbf
		{\bibinfo {volume} {5}},\ \bibinfo {pages} {57} (\bibinfo {year}
		{2014})}\BibitemShut {NoStop}%
	\bibitem [{\citenamefont {Kuramoto}(2008)}]{10.1143/PTPS.176.77}%
	\BibitemOpen
	\bibfield  {author} {\bibinfo {author} {\bibfnamefont {Y.}~\bibnamefont
			{Kuramoto}},\ }\href {\doibase 10.1143/PTPS.176.77} {\bibfield  {journal}
		{\bibinfo  {journal} {Progress of Theoretical Physics Supplement}\ }\textbf
		{\bibinfo {volume} {176}},\ \bibinfo {pages} {77} (\bibinfo {year}
		{2008})}\BibitemShut {NoStop}%
	\bibitem [{\citenamefont {Santini}\ \emph {et~al.}(2009)\citenamefont
		{Santini}, \citenamefont {Carretta}, \citenamefont {Amoretti}, \citenamefont
		{Caciuffo}, \citenamefont {Magnani},\ and\ \citenamefont
		{Lander}}]{Santini2009}%
	\BibitemOpen
	\bibfield  {author} {\bibinfo {author} {\bibfnamefont {P.}~\bibnamefont
			{Santini}}, \bibinfo {author} {\bibfnamefont {S.}~\bibnamefont {Carretta}},
		\bibinfo {author} {\bibfnamefont {G.}~\bibnamefont {Amoretti}}, \bibinfo
		{author} {\bibfnamefont {R.}~\bibnamefont {Caciuffo}}, \bibinfo {author}
		{\bibfnamefont {N.}~\bibnamefont {Magnani}}, \ and\ \bibinfo {author}
		{\bibfnamefont {G.~H.}\ \bibnamefont {Lander}},\ }\href@noop {} {\bibfield
		{journal} {\bibinfo  {journal} {Rev. Mod. Phys.}\ }\textbf {\bibinfo {volume}
			{81}},\ \bibinfo {pages} {807} (\bibinfo {year} {2009})}\BibitemShut
	{NoStop}%
	\bibitem [{\citenamefont {Fu}(2015)}]{PhysRevLett.115.026401}%
	\BibitemOpen
	\bibfield  {author} {\bibinfo {author} {\bibfnamefont {L.}~\bibnamefont
			{Fu}},\ }\href {\doibase 10.1103/PhysRevLett.115.026401} {\bibfield
		{journal} {\bibinfo  {journal} {Phys. Rev. Lett.}\ }\textbf {\bibinfo
			{volume} {115}},\ \bibinfo {pages} {026401} (\bibinfo {year}
		{2015})}\BibitemShut {NoStop}%
	\bibitem [{\citenamefont {Jackeli}\ and\ \citenamefont
		{Khaliullin}(2009)}]{PhysRevLett.103.067205}%
	\BibitemOpen
	\bibfield  {author} {\bibinfo {author} {\bibfnamefont {G.}~\bibnamefont
			{Jackeli}}\ and\ \bibinfo {author} {\bibfnamefont {G.}~\bibnamefont
			{Khaliullin}},\ }\href {\doibase 10.1103/PhysRevLett.103.067205} {\bibfield
		{journal} {\bibinfo  {journal} {Phys. Rev. Lett.}\ }\textbf {\bibinfo
			{volume} {103}},\ \bibinfo {pages} {067205} (\bibinfo {year}
		{2009})}\BibitemShut {NoStop}%
	\bibitem [{\citenamefont {Maharaj}\ \emph {et~al.}(2020)\citenamefont
		{Maharaj}, \citenamefont {Sala}, \citenamefont {Stone}, \citenamefont
		{Kermarrec}, \citenamefont {Ritter}, \citenamefont {Fauth}, \citenamefont
		{Marjerrison}, \citenamefont {Greedan}, \citenamefont {Paramekanti},\ and\
		\citenamefont {Gaulin}}]{Maharaj2020}%
	\BibitemOpen
	\bibfield  {author} {\bibinfo {author} {\bibfnamefont {D.~D.}\ \bibnamefont
			{Maharaj}}, \bibinfo {author} {\bibfnamefont {G.}~\bibnamefont {Sala}},
		\bibinfo {author} {\bibfnamefont {M.~B.}\ \bibnamefont {Stone}}, \bibinfo
		{author} {\bibfnamefont {E.}~\bibnamefont {Kermarrec}}, \bibinfo {author}
		{\bibfnamefont {C.}~\bibnamefont {Ritter}}, \bibinfo {author} {\bibfnamefont
			{F.}~\bibnamefont {Fauth}}, \bibinfo {author} {\bibfnamefont {C.~A.}\
			\bibnamefont {Marjerrison}}, \bibinfo {author} {\bibfnamefont {J.~E.}\
			\bibnamefont {Greedan}}, \bibinfo {author} {\bibfnamefont {A.}~\bibnamefont
			{Paramekanti}}, \ and\ \bibinfo {author} {\bibfnamefont {B.~D.}\ \bibnamefont
			{Gaulin}},\ }\href {\doibase 10.1103/PhysRevLett.124.087206} {\bibfield
		{journal} {\bibinfo  {journal} {Phys. Rev. Lett.}\ }\textbf {\bibinfo
			{volume} {124}},\ \bibinfo {pages} {087206} (\bibinfo {year}
		{2020})}\BibitemShut {NoStop}%
	\bibitem [{\citenamefont {Harter}\ \emph {et~al.}(2017)\citenamefont {Harter},
		\citenamefont {Zhao}, \citenamefont {Yan}, \citenamefont {Mandrus},\ and\
		\citenamefont {Hsieh}}]{doi:10.1126/science.aad1188}%
	\BibitemOpen
	\bibfield  {author} {\bibinfo {author} {\bibfnamefont {J.~W.}\ \bibnamefont
			{Harter}}, \bibinfo {author} {\bibfnamefont {Z.~Y.}\ \bibnamefont {Zhao}},
		\bibinfo {author} {\bibfnamefont {J.-Q.}\ \bibnamefont {Yan}}, \bibinfo
		{author} {\bibfnamefont {D.~G.}\ \bibnamefont {Mandrus}}, \ and\ \bibinfo
		{author} {\bibfnamefont {D.}~\bibnamefont {Hsieh}},\ }\href {\doibase
		10.1126/science.aad1188} {\bibfield  {journal} {\bibinfo  {journal}
			{Science}\ }\textbf {\bibinfo {volume} {356}},\ \bibinfo {pages} {295}
		(\bibinfo {year} {2017})}\BibitemShut {NoStop}%
	\bibitem [{\citenamefont {Khaliullin}\ \emph {et~al.}(2021)\citenamefont
		{Khaliullin}, \citenamefont {Churchill}, \citenamefont {Stavropoulos},\ and\
		\citenamefont {Kee}}]{khaliullin2021}%
	\BibitemOpen
	\bibfield  {author} {\bibinfo {author} {\bibfnamefont {G.}~\bibnamefont
			{Khaliullin}}, \bibinfo {author} {\bibfnamefont {D.}~\bibnamefont
			{Churchill}}, \bibinfo {author} {\bibfnamefont {P.~P.}\ \bibnamefont
			{Stavropoulos}}, \ and\ \bibinfo {author} {\bibfnamefont {H.-Y.}\
			\bibnamefont {Kee}},\ }\href {\doibase 10.1103/PhysRevResearch.3.033163}
	{\bibfield  {journal} {\bibinfo  {journal} {Phys. Rev. Research}\ }\textbf
		{\bibinfo {volume} {3}},\ \bibinfo {pages} {033163} (\bibinfo {year}
		{2021})}\BibitemShut {NoStop}%
	\bibitem [{\citenamefont {Takahashi}\ and\ \citenamefont
		{Shiba}(2000)}]{doi:10.1143/JPSJ.69.3328}%
	\BibitemOpen
	\bibfield  {author} {\bibinfo {author} {\bibfnamefont {A.}~\bibnamefont
			{Takahashi}}\ and\ \bibinfo {author} {\bibfnamefont {H.}~\bibnamefont
			{Shiba}},\ }\href {\doibase 10.1143/JPSJ.69.3328} {\bibfield  {journal}
		{\bibinfo  {journal} {Journal of the Physical Society of Japan}\ }\textbf
		{\bibinfo {volume} {69}},\ \bibinfo {pages} {3328} (\bibinfo {year}
		{2000})}\BibitemShut {NoStop}%
	\bibitem [{\citenamefont {van~den Brink}\ and\ \citenamefont
		{Khomskii}(2001)}]{PhysRevB.63.140416}%
	\BibitemOpen
	\bibfield  {author} {\bibinfo {author} {\bibfnamefont {J.}~\bibnamefont
			{van~den Brink}}\ and\ \bibinfo {author} {\bibfnamefont {D.}~\bibnamefont
			{Khomskii}},\ }\href {\doibase 10.1103/PhysRevB.63.140416} {\bibfield
		{journal} {\bibinfo  {journal} {Phys. Rev. B}\ }\textbf {\bibinfo {volume}
			{63}},\ \bibinfo {pages} {140416} (\bibinfo {year} {2001})}\BibitemShut
	{NoStop}%
	\bibitem [{\citenamefont {Paramekanti}\ \emph {et~al.}(2020)\citenamefont
		{Paramekanti}, \citenamefont {Maharaj},\ and\ \citenamefont
		{Gaulin}}]{Paramekanti2020}%
	\BibitemOpen
	\bibfield  {author} {\bibinfo {author} {\bibfnamefont {A.}~\bibnamefont
			{Paramekanti}}, \bibinfo {author} {\bibfnamefont {D.~D.}\ \bibnamefont
			{Maharaj}}, \ and\ \bibinfo {author} {\bibfnamefont {B.~D.}\ \bibnamefont
			{Gaulin}},\ }\href {\doibase 10.1103/PhysRevB.101.054439} {\bibfield
		{journal} {\bibinfo  {journal} {Phys. Rev. B}\ }\textbf {\bibinfo {volume}
			{101}},\ \bibinfo {pages} {054439} (\bibinfo {year} {2020})}\BibitemShut
	{NoStop}%
	\bibitem [{\citenamefont {Voleti}\ \emph {et~al.}(2020)\citenamefont {Voleti},
		\citenamefont {Maharaj}, \citenamefont {Gaulin}, \citenamefont {Luke},\ and\
		\citenamefont {Paramekanti}}]{PhysRevB.101.155118}%
	\BibitemOpen
	\bibfield  {author} {\bibinfo {author} {\bibfnamefont {S.}~\bibnamefont
			{Voleti}}, \bibinfo {author} {\bibfnamefont {D.~D.}\ \bibnamefont {Maharaj}},
		\bibinfo {author} {\bibfnamefont {B.~D.}\ \bibnamefont {Gaulin}}, \bibinfo
		{author} {\bibfnamefont {G.}~\bibnamefont {Luke}}, \ and\ \bibinfo {author}
		{\bibfnamefont {A.}~\bibnamefont {Paramekanti}},\ }\href {\doibase
		10.1103/PhysRevB.101.155118} {\bibfield  {journal} {\bibinfo  {journal}
			{Phys. Rev. B}\ }\textbf {\bibinfo {volume} {101}},\ \bibinfo {pages}
		{155118} (\bibinfo {year} {2020})}\BibitemShut {NoStop}%
	\bibitem [{\citenamefont {Lovesey}\ and\ \citenamefont
		{Khalyavin}(2020{\natexlab{a}})}]{PhysRevB.102.064407}%
	\BibitemOpen
	\bibfield  {author} {\bibinfo {author} {\bibfnamefont {S.~W.}\ \bibnamefont
			{Lovesey}}\ and\ \bibinfo {author} {\bibfnamefont {D.~D.}\ \bibnamefont
			{Khalyavin}},\ }\href {\doibase 10.1103/PhysRevB.102.064407} {\bibfield
		{journal} {\bibinfo  {journal} {Phys. Rev. B}\ }\textbf {\bibinfo {volume}
			{102}},\ \bibinfo {pages} {064407} (\bibinfo {year}
		{2020}{\natexlab{a}})}\BibitemShut {NoStop}%
	\bibitem [{\citenamefont {Chen}\ and\ \citenamefont
		{Balents}(2011)}]{PhysRevB.84.094420}%
	\BibitemOpen
	\bibfield  {author} {\bibinfo {author} {\bibfnamefont {G.}~\bibnamefont
			{Chen}}\ and\ \bibinfo {author} {\bibfnamefont {L.}~\bibnamefont {Balents}},\
	}\href {\doibase 10.1103/PhysRevB.84.094420} {\bibfield  {journal} {\bibinfo
			{journal} {Phys. Rev. B}\ }\textbf {\bibinfo {volume} {84}},\ \bibinfo
		{pages} {094420} (\bibinfo {year} {2011})}\BibitemShut {NoStop}%
	\bibitem [{\citenamefont {Yamamura}\ \emph {et~al.}(2006)\citenamefont
		{Yamamura}, \citenamefont {Wakeshima},\ and\ \citenamefont
		{Hinatsu}}]{Yamamura2006}%
	\BibitemOpen
	\bibfield  {author} {\bibinfo {author} {\bibfnamefont {K.}~\bibnamefont
			{Yamamura}}, \bibinfo {author} {\bibfnamefont {M.}~\bibnamefont {Wakeshima}},
		\ and\ \bibinfo {author} {\bibfnamefont {Y.}~\bibnamefont {Hinatsu}},\ }\href
	{\doibase https://doi.org/10.1016/j.jssc.2005.10.003} {\bibfield  {journal}
		{\bibinfo  {journal} {Journal of Solid State Chemistry}\ }\textbf {\bibinfo
			{volume} {179}},\ \bibinfo {pages} {605} (\bibinfo {year}
		{2006})}\BibitemShut {NoStop}%
	\bibitem [{\citenamefont {Thompson}\ \emph {et~al.}(2014)\citenamefont
		{Thompson}, \citenamefont {Carlo}, \citenamefont {Flacau}, \citenamefont
		{Aharen}, \citenamefont {Leahy}, \citenamefont {Pollichemi}, \citenamefont
		{Munsie}, \citenamefont {Medina}, \citenamefont {Luke}, \citenamefont
		{Munevar}, \citenamefont {Cheung}, \citenamefont {Goko}, \citenamefont
		{Uemura},\ and\ \citenamefont {Greedan}}]{Thompson2014}%
	\BibitemOpen
	\bibfield  {author} {\bibinfo {author} {\bibfnamefont {C.~M.}\ \bibnamefont
			{Thompson}}, \bibinfo {author} {\bibfnamefont {J.~P.}\ \bibnamefont {Carlo}},
		\bibinfo {author} {\bibfnamefont {R.}~\bibnamefont {Flacau}}, \bibinfo
		{author} {\bibfnamefont {T.}~\bibnamefont {Aharen}}, \bibinfo {author}
		{\bibfnamefont {I.~A.}\ \bibnamefont {Leahy}}, \bibinfo {author}
		{\bibfnamefont {J.~R.}\ \bibnamefont {Pollichemi}}, \bibinfo {author}
		{\bibfnamefont {T.~J.~S.}\ \bibnamefont {Munsie}}, \bibinfo {author}
		{\bibfnamefont {T.}~\bibnamefont {Medina}}, \bibinfo {author} {\bibfnamefont
			{G.~M.}\ \bibnamefont {Luke}}, \bibinfo {author} {\bibfnamefont
			{J.}~\bibnamefont {Munevar}}, \bibinfo {author} {\bibfnamefont
			{S.}~\bibnamefont {Cheung}}, \bibinfo {author} {\bibfnamefont
			{T.}~\bibnamefont {Goko}}, \bibinfo {author} {\bibfnamefont {Y.~J.}\
			\bibnamefont {Uemura}}, \ and\ \bibinfo {author} {\bibfnamefont {J.~E.}\
			\bibnamefont {Greedan}},\ }\href {\doibase 10.1088/0953-8984/26/30/306003}
	{\bibfield  {journal} {\bibinfo  {journal} {Journal of Physics: Condensed
				Matter}\ }\textbf {\bibinfo {volume} {26}},\ \bibinfo {pages} {306003}
		(\bibinfo {year} {2014})}\BibitemShut {NoStop}%
	\bibitem [{\citenamefont {Marjerrison}\ \emph {et~al.}(2016)\citenamefont
		{Marjerrison}, \citenamefont {Thompson}, \citenamefont {Sharma},
		\citenamefont {Hallas}, \citenamefont {Wilson}, \citenamefont {Munsie},
		\citenamefont {Flacau}, \citenamefont {Wiebe}, \citenamefont {Gaulin},
		\citenamefont {Luke},\ and\ \citenamefont {Greedan}}]{Marjerrison2016}%
	\BibitemOpen
	\bibfield  {author} {\bibinfo {author} {\bibfnamefont {C.~A.}\ \bibnamefont
			{Marjerrison}}, \bibinfo {author} {\bibfnamefont {C.~M.}\ \bibnamefont
			{Thompson}}, \bibinfo {author} {\bibfnamefont {A.~Z.}\ \bibnamefont
			{Sharma}}, \bibinfo {author} {\bibfnamefont {A.~M.}\ \bibnamefont {Hallas}},
		\bibinfo {author} {\bibfnamefont {M.~N.}\ \bibnamefont {Wilson}}, \bibinfo
		{author} {\bibfnamefont {T.~J.~S.}\ \bibnamefont {Munsie}}, \bibinfo {author}
		{\bibfnamefont {R.}~\bibnamefont {Flacau}}, \bibinfo {author} {\bibfnamefont
			{C.~R.}\ \bibnamefont {Wiebe}}, \bibinfo {author} {\bibfnamefont {B.~D.}\
			\bibnamefont {Gaulin}}, \bibinfo {author} {\bibfnamefont {G.~M.}\
			\bibnamefont {Luke}}, \ and\ \bibinfo {author} {\bibfnamefont {J.~E.}\
			\bibnamefont {Greedan}},\ }\href {\doibase 10.1103/PhysRevB.94.134429}
	{\bibfield  {journal} {\bibinfo  {journal} {Phys. Rev. B}\ }\textbf {\bibinfo
			{volume} {94}},\ \bibinfo {pages} {134429} (\bibinfo {year}
		{2016})}\BibitemShut {NoStop}%
	\bibitem [{\citenamefont {Churchill}\ and\ \citenamefont
		{Kee}()}]{Churchill2021}%
	\BibitemOpen
	\bibfield  {author} {\bibinfo {author} {\bibfnamefont {D.}~\bibnamefont
			{Churchill}}\ and\ \bibinfo {author} {\bibfnamefont {H.-Y.}\ \bibnamefont
			{Kee}},\ }\href@noop {} {\bibinfo  {journal} {arXiv:2109.08104
			[cond-mat.str-el]}\ }\BibitemShut {NoStop}%
	\bibitem [{\citenamefont {Voleti}\ \emph {et~al.}()\citenamefont {Voleti},
		\citenamefont {Haldar},\ and\ \citenamefont {Paramekanti}}]{Voleti2021}%
	\BibitemOpen
	\bibfield  {journal} {  }\bibfield  {author} {\bibinfo {author} {\bibfnamefont
			{S.}~\bibnamefont {Voleti}}, \bibinfo {author} {\bibfnamefont
			{A.}~\bibnamefont {Haldar}}, \ and\ \bibinfo {author} {\bibfnamefont
			{A.}~\bibnamefont {Paramekanti}},\ }\href@noop {} {\bibinfo  {journal}
		{arXiv:2109.13266 [cond-mat.str-el]}\ }\BibitemShut {NoStop}%
	\bibitem [{\citenamefont {Blaha}\ \emph {et~al.}(2018)\citenamefont {Blaha},
		\citenamefont {Schwarz}, \citenamefont {Madsen}, \citenamefont {Kvasnicka},
		\citenamefont {Luitz}, \citenamefont {Laskowski}, \citenamefont {Tran},\ and\
		\citenamefont {Marks}}]{Wien2k}%
	\BibitemOpen
	\bibfield  {journal} {  }\bibfield  {author} {\bibinfo {author} {\bibfnamefont
			{P.}~\bibnamefont {Blaha}}, \bibinfo {author} {\bibfnamefont
			{K.}~\bibnamefont {Schwarz}}, \bibinfo {author} {\bibfnamefont
			{G.}~\bibnamefont {Madsen}}, \bibinfo {author} {\bibfnamefont
			{D.}~\bibnamefont {Kvasnicka}}, \bibinfo {author} {\bibfnamefont
			{J.}~\bibnamefont {Luitz}}, \bibinfo {author} {\bibfnamefont
			{R.}~\bibnamefont {Laskowski}}, \bibinfo {author} {\bibfnamefont
			{F.}~\bibnamefont {Tran}}, \ and\ \bibinfo {author} {\bibfnamefont {L.~D.}\
			\bibnamefont {Marks}},\ }\href@noop {} {\emph {\bibinfo {title} {WIEN2k, An
				augmented Plane Wave + Local Orbitals Program for Calculating Crystal
				Properties}}}\ (\bibinfo  {publisher} {Karlheinz Schwarz, Techn. Universität
		Wien, Austria,ISBN 3-9501031-1-2},\ \bibinfo {year} {2018})\BibitemShut
	{NoStop}%
	\bibitem [{\citenamefont {Georges}\ \emph {et~al.}(1996)\citenamefont
		{Georges}, \citenamefont {Kotliar}, \citenamefont {Krauth},\ and\
		\citenamefont {Rozenberg}}]{Georges1996}%
	\BibitemOpen
	\bibfield  {author} {\bibinfo {author} {\bibfnamefont {A.}~\bibnamefont
			{Georges}}, \bibinfo {author} {\bibfnamefont {G.}~\bibnamefont {Kotliar}},
		\bibinfo {author} {\bibfnamefont {W.}~\bibnamefont {Krauth}}, \ and\ \bibinfo
		{author} {\bibfnamefont {M.~J.}\ \bibnamefont {Rozenberg}},\ }\href@noop {}
	{\bibfield  {journal} {\bibinfo  {journal} {Rev. Mod. Phys.}\ }\textbf
		{\bibinfo {volume} {68}},\ \bibinfo {pages} {13} (\bibinfo {year}
		{1996})}\BibitemShut {NoStop}%
	\bibitem [{\citenamefont {Anisimov}\ \emph {et~al.}(1997)\citenamefont
		{Anisimov}, \citenamefont {Poteryaev}, \citenamefont {Korotin}, \citenamefont
		{Anokhin},\ and\ \citenamefont {Kotliar}}]{Anisimov1997_1}%
	\BibitemOpen
	\bibfield  {author} {\bibinfo {author} {\bibfnamefont {V.~I.}\ \bibnamefont
			{Anisimov}}, \bibinfo {author} {\bibfnamefont {A.~I.}\ \bibnamefont
			{Poteryaev}}, \bibinfo {author} {\bibfnamefont {M.~A.}\ \bibnamefont
			{Korotin}}, \bibinfo {author} {\bibfnamefont {A.~O.}\ \bibnamefont
			{Anokhin}}, \ and\ \bibinfo {author} {\bibfnamefont {G.}~\bibnamefont
			{Kotliar}},\ }\href@noop {} {\bibfield  {journal} {\bibinfo  {journal}
			{Journal of Physics: Condensed Matter}\ }\textbf {\bibinfo {volume} {9}},\
		\bibinfo {pages} {7359} (\bibinfo {year} {1997})}\BibitemShut {NoStop}%
	\bibitem [{\citenamefont {Lichtenstein}\ and\ \citenamefont
		{Katsnelson}(1998)}]{Lichtenstein_LDApp}%
	\BibitemOpen
	\bibfield  {author} {\bibinfo {author} {\bibfnamefont {A.~I.}\ \bibnamefont
			{Lichtenstein}}\ and\ \bibinfo {author} {\bibfnamefont {M.~I.}\ \bibnamefont
			{Katsnelson}},\ }\href@noop {} {\bibfield  {journal} {\bibinfo  {journal}
			{Phys. Rev. B}\ }\textbf {\bibinfo {volume} {57}},\ \bibinfo {pages} {6884}
		(\bibinfo {year} {1998})}\BibitemShut {NoStop}%
	\bibitem [{\citenamefont {Aichhorn}\ \emph {et~al.}(2016)\citenamefont
		{Aichhorn}, \citenamefont {Pourovskii}, \citenamefont {Seth}, \citenamefont
		{Vildosola}, \citenamefont {Zingl}, \citenamefont {Peil}, \citenamefont
		{Deng}, \citenamefont {Mravlje}, \citenamefont {Kraberger}, \citenamefont
		{Martins} \emph {et~al.}}]{Aichhorn2016}%
	\BibitemOpen
	\bibfield  {author} {\bibinfo {author} {\bibfnamefont {M.}~\bibnamefont
			{Aichhorn}}, \bibinfo {author} {\bibfnamefont {L.}~\bibnamefont
			{Pourovskii}}, \bibinfo {author} {\bibfnamefont {P.}~\bibnamefont {Seth}},
		\bibinfo {author} {\bibfnamefont {V.}~\bibnamefont {Vildosola}}, \bibinfo
		{author} {\bibfnamefont {M.}~\bibnamefont {Zingl}}, \bibinfo {author}
		{\bibfnamefont {O.~E.}\ \bibnamefont {Peil}}, \bibinfo {author}
		{\bibfnamefont {X.}~\bibnamefont {Deng}}, \bibinfo {author} {\bibfnamefont
			{J.}~\bibnamefont {Mravlje}}, \bibinfo {author} {\bibfnamefont {G.~J.}\
			\bibnamefont {Kraberger}}, \bibinfo {author} {\bibfnamefont {C.}~\bibnamefont
			{Martins}},  \emph {et~al.},\ }\href@noop {} {\bibfield  {journal} {\bibinfo
			{journal} {Computer Physics Communications}\ }\textbf {\bibinfo {volume}
			{204}},\ \bibinfo {pages} {200} (\bibinfo {year} {2016})}\BibitemShut
	{NoStop}%
	\bibitem [{\citenamefont {Hubbard}(1963)}]{hubbard_1}%
	\BibitemOpen
	\bibfield  {author} {\bibinfo {author} {\bibfnamefont {J.}~\bibnamefont
			{Hubbard}},\ }\href@noop {} {\bibfield  {journal} {\bibinfo  {journal} {Proc.
				Roy. Soc. (London)}\ }\textbf {\bibinfo {volume} {A 276}},\ \bibinfo {pages}
		{238} (\bibinfo {year} {1963})}\BibitemShut {NoStop}%
	\bibitem [{\citenamefont {Pourovskii}(2016)}]{Pourovskii2016}%
	\BibitemOpen
	\bibfield  {author} {\bibinfo {author} {\bibfnamefont {L.~V.}\ \bibnamefont
			{Pourovskii}},\ }\href@noop {} {\bibfield  {journal} {\bibinfo  {journal}
			{Phys. Rev. B}\ }\textbf {\bibinfo {volume} {94}},\ \bibinfo {pages} {115117}
		(\bibinfo {year} {2016})}\BibitemShut {NoStop}%
	\bibitem [{sup()}]{supplmat}%
	\BibitemOpen
	\href@noop {} {}\bibinfo {note} {See Supplementary Material [url], which
		includes
		Refs.~\cite{Aichhorn2009,Aichhorn2011,Anisimov1993,FioreMosca2021,Erickson2007,Georges_Hunds_materials,Winter2016,Taylor2017,Yan2017,Kobayashi2011,Stassis1976,Pourovskii2020,Amadon2008}}\BibitemShut
	{NoStop}%
	\bibitem [{\citenamefont {Delange}\ \emph {et~al.}(2017)\citenamefont
		{Delange}, \citenamefont {Biermann}, \citenamefont {Miyake},\ and\
		\citenamefont {Pourovskii}}]{Delange2017}%
	\BibitemOpen
	\bibfield  {author} {\bibinfo {author} {\bibfnamefont {P.}~\bibnamefont
			{Delange}}, \bibinfo {author} {\bibfnamefont {S.}~\bibnamefont {Biermann}},
		\bibinfo {author} {\bibfnamefont {T.}~\bibnamefont {Miyake}}, \ and\ \bibinfo
		{author} {\bibfnamefont {L.}~\bibnamefont {Pourovskii}},\ }\href {\doibase
		10.1103/PhysRevB.96.155132} {\bibfield  {journal} {\bibinfo  {journal} {Phys.
				Rev. B}\ }\textbf {\bibinfo {volume} {96}},\ \bibinfo {pages} {155132}
		(\bibinfo {year} {2017})}\BibitemShut {NoStop}%
	\bibitem [{\citenamefont {Yamada}\ \emph {et~al.}(2017)\citenamefont {Yamada},
		\citenamefont {Takamatsu}, \citenamefont {Hayashi},\ and\ \citenamefont
		{Ikeno}}]{Yamada2017}%
	\BibitemOpen
	\bibfield  {author} {\bibinfo {author} {\bibfnamefont {I.}~\bibnamefont
			{Yamada}}, \bibinfo {author} {\bibfnamefont {A.}~\bibnamefont {Takamatsu}},
		\bibinfo {author} {\bibfnamefont {N.}~\bibnamefont {Hayashi}}, \ and\
		\bibinfo {author} {\bibfnamefont {H.}~\bibnamefont {Ikeno}},\ }\href
	{\doibase 10.1021/acs.inorgchem.7b01405} {\bibfield  {journal} {\bibinfo
			{journal} {Inorganic Chemistry}\ }\textbf {\bibinfo {volume} {56}},\ \bibinfo
		{pages} {9303} (\bibinfo {year} {2017})}\BibitemShut {NoStop}%
	\bibitem [{\citenamefont {Rotter}(2004)}]{Rotter2004}%
	\BibitemOpen
	\bibfield  {author} {\bibinfo {author} {\bibfnamefont {M.}~\bibnamefont
			{Rotter}},\ }\href {http://www.mcphase.de/} {\bibfield  {journal} {\bibinfo
			{journal} {Journal of Magnetism and Magnetic Materials}\ }\textbf {\bibinfo
			{volume} {272-276, Supplement}},\ \bibinfo {pages} {E481 } (\bibinfo {year}
		{2004})}\BibitemShut {NoStop}%
	\bibitem [{\citenamefont {Horvat}\ \emph {et~al.}(2017)\citenamefont {Horvat},
		\citenamefont {Pourovskii}, \citenamefont {Aichhorn},\ and\ \citenamefont
		{Mravlje}}]{Horvat2017}%
	\BibitemOpen
	\bibfield  {author} {\bibinfo {author} {\bibfnamefont {A.}~\bibnamefont
			{Horvat}}, \bibinfo {author} {\bibfnamefont {L.}~\bibnamefont {Pourovskii}},
		\bibinfo {author} {\bibfnamefont {M.}~\bibnamefont {Aichhorn}}, \ and\
		\bibinfo {author} {\bibfnamefont {J.}~\bibnamefont {Mravlje}},\ }\href
	{\doibase 10.1103/PhysRevB.95.205115} {\bibfield  {journal} {\bibinfo
			{journal} {Phys. Rev. B}\ }\textbf {\bibinfo {volume} {95}},\ \bibinfo
		{pages} {205115} (\bibinfo {year} {2017})}\BibitemShut {NoStop}%
	\bibitem [{\citenamefont {Pourovskii}\ and\ \citenamefont
		{Khmelevskyi}(2019)}]{Pourovskii2019}%
	\BibitemOpen
	\bibfield  {author} {\bibinfo {author} {\bibfnamefont {L.~V.}\ \bibnamefont
			{Pourovskii}}\ and\ \bibinfo {author} {\bibfnamefont {S.}~\bibnamefont
			{Khmelevskyi}},\ }\href {\doibase 10.1103/PhysRevB.99.094439} {\bibfield
		{journal} {\bibinfo  {journal} {Phys. Rev. B}\ }\textbf {\bibinfo {volume}
			{99}},\ \bibinfo {pages} {094439} (\bibinfo {year} {2019})}\BibitemShut
	{NoStop}%
	\bibitem [{\citenamefont {Pourovskii}\ and\ \citenamefont
		{Khmelevskyi}(2021)}]{Pourovskii2021}%
	\BibitemOpen
	\bibfield  {author} {\bibinfo {author} {\bibfnamefont {L.~V.}\ \bibnamefont
			{Pourovskii}}\ and\ \bibinfo {author} {\bibfnamefont {S.}~\bibnamefont
			{Khmelevskyi}},\ }\href {\doibase 10.1073/pnas.2025317118} {\bibfield
		{journal} {\bibinfo  {journal} {Proceedings of the National Academy of
				Sciences}\ }\textbf {\bibinfo {volume} {118}},\ \bibinfo {pages}
		{e2025317118} (\bibinfo {year} {2021})}\BibitemShut {NoStop}%
	\bibitem [{\citenamefont {Jensen}\ and\ \citenamefont
		{Mackintosh}(1991)}]{RareEarthMag_book}%
	\BibitemOpen
	\bibfield  {author} {\bibinfo {author} {\bibfnamefont {J.}~\bibnamefont
			{Jensen}}\ and\ \bibinfo {author} {\bibfnamefont {A.~R.}\ \bibnamefont
			{Mackintosh}},\ }\href@noop {} {\emph {\bibinfo {title} {Rare Earth
				Magnetism: Structures and Excitations}}}\ (\bibinfo  {publisher} {Clarendon
		Press, Oxford},\ \bibinfo {year} {1991})\BibitemShut {NoStop}%
	\bibitem [{\citenamefont {Shiina}\ \emph {et~al.}(2007)\citenamefont {Shiina},
		\citenamefont {Sakai},\ and\ \citenamefont {Shiba}}]{Shiina2007}%
	\BibitemOpen
	\bibfield  {author} {\bibinfo {author} {\bibfnamefont {R.}~\bibnamefont
			{Shiina}}, \bibinfo {author} {\bibfnamefont {O.}~\bibnamefont {Sakai}}, \
		and\ \bibinfo {author} {\bibfnamefont {H.}~\bibnamefont {Shiba}},\ }\href
	{\doibase 10.1143/JPSJ.76.094702} {\bibfield  {journal} {\bibinfo  {journal}
			{Journal of the Physical Society of Japan}\ }\textbf {\bibinfo {volume}
			{76}},\ \bibinfo {pages} {094702} (\bibinfo {year} {2007})}\BibitemShut
	{NoStop}%
	\bibitem [{\citenamefont {Lovesey}(1984)}]{Lovesey_book_full}%
	\BibitemOpen
	\bibfield  {author} {\bibinfo {author} {\bibfnamefont {S.~W.}\ \bibnamefont
			{Lovesey}},\ }\href@noop {} {\emph {\bibinfo {title} {Theory of Neutron
				Scattering from Condensed Matter}}}\ (\bibinfo  {publisher} {Clarendon Press,
		Oxford},\ \bibinfo {year} {1984})\BibitemShut {NoStop}%
	\bibitem [{\citenamefont {Lovesey}\ and\ \citenamefont
		{Khalyavin}(2020{\natexlab{b}})}]{Lovesey2020}%
	\BibitemOpen
	\bibfield  {author} {\bibinfo {author} {\bibfnamefont {S.~W.}\ \bibnamefont
			{Lovesey}}\ and\ \bibinfo {author} {\bibfnamefont {D.~D.}\ \bibnamefont
			{Khalyavin}},\ }\href {\doibase 10.1103/PhysRevB.102.064407} {\bibfield
		{journal} {\bibinfo  {journal} {Phys. Rev. B}\ }\textbf {\bibinfo {volume}
			{102}},\ \bibinfo {pages} {064407} (\bibinfo {year}
		{2020}{\natexlab{b}})}\BibitemShut {NoStop}%
	\bibitem [{\citenamefont {Aichhorn}\ \emph {et~al.}(2009)\citenamefont
		{Aichhorn}, \citenamefont {Pourovskii}, \citenamefont {Vildosola},
		\citenamefont {Ferrero}, \citenamefont {Parcollet}, \citenamefont {Miyake},
		\citenamefont {Georges},\ and\ \citenamefont {Biermann}}]{Aichhorn2009}%
	\BibitemOpen
	\bibfield  {author} {\bibinfo {author} {\bibfnamefont {M.}~\bibnamefont
			{Aichhorn}}, \bibinfo {author} {\bibfnamefont {L.}~\bibnamefont
			{Pourovskii}}, \bibinfo {author} {\bibfnamefont {V.}~\bibnamefont
			{Vildosola}}, \bibinfo {author} {\bibfnamefont {M.}~\bibnamefont {Ferrero}},
		\bibinfo {author} {\bibfnamefont {O.}~\bibnamefont {Parcollet}}, \bibinfo
		{author} {\bibfnamefont {T.}~\bibnamefont {Miyake}}, \bibinfo {author}
		{\bibfnamefont {A.}~\bibnamefont {Georges}}, \ and\ \bibinfo {author}
		{\bibfnamefont {S.}~\bibnamefont {Biermann}},\ }\href@noop {} {\bibfield
		{journal} {\bibinfo  {journal} {Phys. Rev. B}\ }\textbf {\bibinfo {volume}
			{80}},\ \bibinfo {pages} {085101} (\bibinfo {year} {2009})}\BibitemShut
	{NoStop}%
	\bibitem [{\citenamefont {Aichhorn}\ \emph {et~al.}(2011)\citenamefont
		{Aichhorn}, \citenamefont {Pourovskii},\ and\ \citenamefont
		{Georges}}]{Aichhorn2011}%
	\BibitemOpen
	\bibfield  {author} {\bibinfo {author} {\bibfnamefont {M.}~\bibnamefont
			{Aichhorn}}, \bibinfo {author} {\bibfnamefont {L.}~\bibnamefont
			{Pourovskii}}, \ and\ \bibinfo {author} {\bibfnamefont {A.}~\bibnamefont
			{Georges}},\ }\href@noop {} {\bibfield  {journal} {\bibinfo  {journal} {Phys.
				Rev. B}\ }\textbf {\bibinfo {volume} {84}},\ \bibinfo {pages} {054529}
		(\bibinfo {year} {2011})}\BibitemShut {NoStop}%
	\bibitem [{\citenamefont {Anisimov}\ \emph {et~al.}(1993)\citenamefont
		{Anisimov}, \citenamefont {Solovyev}, \citenamefont {Korotin}, \citenamefont
		{Czy\ifmmode~\dot{z}\else \.{z}\fi{}yk},\ and\ \citenamefont
		{Sawatzky}}]{Anisimov1993}%
	\BibitemOpen
	\bibfield  {author} {\bibinfo {author} {\bibfnamefont {V.~I.}\ \bibnamefont
			{Anisimov}}, \bibinfo {author} {\bibfnamefont {I.~V.}\ \bibnamefont
			{Solovyev}}, \bibinfo {author} {\bibfnamefont {M.~A.}\ \bibnamefont
			{Korotin}}, \bibinfo {author} {\bibfnamefont {M.~T.}\ \bibnamefont
			{Czy\ifmmode~\dot{z}\else \.{z}\fi{}yk}}, \ and\ \bibinfo {author}
		{\bibfnamefont {G.~A.}\ \bibnamefont {Sawatzky}},\ }\href {\doibase
		10.1103/PhysRevB.48.16929} {\bibfield  {journal} {\bibinfo  {journal} {Phys.
				Rev. B}\ }\textbf {\bibinfo {volume} {48}},\ \bibinfo {pages} {16929}
		(\bibinfo {year} {1993})}\BibitemShut {NoStop}%
	\bibitem [{\citenamefont {Fiore~Mosca}\ \emph {et~al.}(2021)\citenamefont
		{Fiore~Mosca}, \citenamefont {Pourovskii}, \citenamefont {Kim}, \citenamefont
		{Liu}, \citenamefont {Sanna}, \citenamefont {Boscherini}, \citenamefont
		{Khmelevskyi},\ and\ \citenamefont {Franchini}}]{FioreMosca2021}%
	\BibitemOpen
	\bibfield  {author} {\bibinfo {author} {\bibfnamefont {D.}~\bibnamefont
			{Fiore~Mosca}}, \bibinfo {author} {\bibfnamefont {L.~V.}\ \bibnamefont
			{Pourovskii}}, \bibinfo {author} {\bibfnamefont {B.~H.}\ \bibnamefont {Kim}},
		\bibinfo {author} {\bibfnamefont {P.}~\bibnamefont {Liu}}, \bibinfo {author}
		{\bibfnamefont {S.}~\bibnamefont {Sanna}}, \bibinfo {author} {\bibfnamefont
			{F.}~\bibnamefont {Boscherini}}, \bibinfo {author} {\bibfnamefont
			{S.}~\bibnamefont {Khmelevskyi}}, \ and\ \bibinfo {author} {\bibfnamefont
			{C.}~\bibnamefont {Franchini}},\ }\href {\doibase
		10.1103/PhysRevB.103.104401} {\bibfield  {journal} {\bibinfo  {journal}
			{Phys. Rev. B}\ }\textbf {\bibinfo {volume} {103}},\ \bibinfo {pages}
		{104401} (\bibinfo {year} {2021})}\BibitemShut {NoStop}%
	\bibitem [{\citenamefont {Erickson}\ \emph {et~al.}(2007)\citenamefont
		{Erickson}, \citenamefont {Misra}, \citenamefont {Miller}, \citenamefont
		{Gupta}, \citenamefont {Schlesinger}, \citenamefont {Harrison}, \citenamefont
		{Kim},\ and\ \citenamefont {Fisher}}]{Erickson2007}%
	\BibitemOpen
	\bibfield  {author} {\bibinfo {author} {\bibfnamefont {A.~S.}\ \bibnamefont
			{Erickson}}, \bibinfo {author} {\bibfnamefont {S.}~\bibnamefont {Misra}},
		\bibinfo {author} {\bibfnamefont {G.~J.}\ \bibnamefont {Miller}}, \bibinfo
		{author} {\bibfnamefont {R.~R.}\ \bibnamefont {Gupta}}, \bibinfo {author}
		{\bibfnamefont {Z.}~\bibnamefont {Schlesinger}}, \bibinfo {author}
		{\bibfnamefont {W.~A.}\ \bibnamefont {Harrison}}, \bibinfo {author}
		{\bibfnamefont {J.~M.}\ \bibnamefont {Kim}}, \ and\ \bibinfo {author}
		{\bibfnamefont {I.~R.}\ \bibnamefont {Fisher}},\ }\href {\doibase
		10.1103/PhysRevLett.99.016404} {\bibfield  {journal} {\bibinfo  {journal}
			{Phys. Rev. Lett.}\ }\textbf {\bibinfo {volume} {99}},\ \bibinfo {pages}
		{016404} (\bibinfo {year} {2007})}\BibitemShut {NoStop}%
	\bibitem [{\citenamefont {Georges}\ \emph {et~al.}(2013)\citenamefont
		{Georges}, \citenamefont {Medici},\ and\ \citenamefont
		{Mravlje}}]{Georges_Hunds_materials}%
	\BibitemOpen
	\bibfield  {author} {\bibinfo {author} {\bibfnamefont {A.}~\bibnamefont
			{Georges}}, \bibinfo {author} {\bibfnamefont {L.~d.}\ \bibnamefont {Medici}},
		\ and\ \bibinfo {author} {\bibfnamefont {J.}~\bibnamefont {Mravlje}},\ }\href
	{\doibase 10.1146/annurev-conmatphys-020911-125045} {\bibfield  {journal}
		{\bibinfo  {journal} {Annual Review of Condensed Matter Physics}\ }\textbf
		{\bibinfo {volume} {4}},\ \bibinfo {pages} {137} (\bibinfo {year}
		{2013})}\BibitemShut {NoStop}%
	\bibitem [{\citenamefont {Winter}\ \emph {et~al.}(2016)\citenamefont {Winter},
		\citenamefont {Li}, \citenamefont {Jeschke},\ and\ \citenamefont
		{Valent\'{\i}}}]{Winter2016}%
	\BibitemOpen
	\bibfield  {author} {\bibinfo {author} {\bibfnamefont {S.~M.}\ \bibnamefont
			{Winter}}, \bibinfo {author} {\bibfnamefont {Y.}~\bibnamefont {Li}}, \bibinfo
		{author} {\bibfnamefont {H.~O.}\ \bibnamefont {Jeschke}}, \ and\ \bibinfo
		{author} {\bibfnamefont {R.}~\bibnamefont {Valent\'{\i}}},\ }\href {\doibase
		10.1103/PhysRevB.93.214431} {\bibfield  {journal} {\bibinfo  {journal} {Phys.
				Rev. B}\ }\textbf {\bibinfo {volume} {93}},\ \bibinfo {pages} {214431}
		(\bibinfo {year} {2016})}\BibitemShut {NoStop}%
	\bibitem [{\citenamefont {Taylor}\ \emph {et~al.}(2017)\citenamefont {Taylor},
		\citenamefont {Calder}, \citenamefont {Morrow}, \citenamefont {Feng},
		\citenamefont {Upton}, \citenamefont {Lumsden}, \citenamefont {Yamaura},
		\citenamefont {Woodward},\ and\ \citenamefont {Christianson}}]{Taylor2017}%
	\BibitemOpen
	\bibfield  {author} {\bibinfo {author} {\bibfnamefont {A.~E.}\ \bibnamefont
			{Taylor}}, \bibinfo {author} {\bibfnamefont {S.}~\bibnamefont {Calder}},
		\bibinfo {author} {\bibfnamefont {R.}~\bibnamefont {Morrow}}, \bibinfo
		{author} {\bibfnamefont {H.~L.}\ \bibnamefont {Feng}}, \bibinfo {author}
		{\bibfnamefont {M.~H.}\ \bibnamefont {Upton}}, \bibinfo {author}
		{\bibfnamefont {M.~D.}\ \bibnamefont {Lumsden}}, \bibinfo {author}
		{\bibfnamefont {K.}~\bibnamefont {Yamaura}}, \bibinfo {author} {\bibfnamefont
			{P.~M.}\ \bibnamefont {Woodward}}, \ and\ \bibinfo {author} {\bibfnamefont
			{A.~D.}\ \bibnamefont {Christianson}},\ }\href {\doibase
		10.1103/PhysRevLett.118.207202} {\bibfield  {journal} {\bibinfo  {journal}
			{Phys. Rev. Lett.}\ }\textbf {\bibinfo {volume} {118}},\ \bibinfo {pages}
		{207202} (\bibinfo {year} {2017})}\BibitemShut {NoStop}%
	\bibitem [{\citenamefont {Yuan}\ \emph {et~al.}(2017)\citenamefont {Yuan},
		\citenamefont {Clancy}, \citenamefont {Cook}, \citenamefont {Thompson},
		\citenamefont {Greedan}, \citenamefont {Cao}, \citenamefont {Jeon},
		\citenamefont {Noh}, \citenamefont {Upton}, \citenamefont {Casa},
		\citenamefont {Gog}, \citenamefont {Paramekanti},\ and\ \citenamefont
		{Kim}}]{Yan2017}%
	\BibitemOpen
	\bibfield  {author} {\bibinfo {author} {\bibfnamefont {B.}~\bibnamefont
			{Yuan}}, \bibinfo {author} {\bibfnamefont {J.~P.}\ \bibnamefont {Clancy}},
		\bibinfo {author} {\bibfnamefont {A.~M.}\ \bibnamefont {Cook}}, \bibinfo
		{author} {\bibfnamefont {C.~M.}\ \bibnamefont {Thompson}}, \bibinfo {author}
		{\bibfnamefont {J.}~\bibnamefont {Greedan}}, \bibinfo {author} {\bibfnamefont
			{G.}~\bibnamefont {Cao}}, \bibinfo {author} {\bibfnamefont {B.~C.}\
			\bibnamefont {Jeon}}, \bibinfo {author} {\bibfnamefont {T.~W.}\ \bibnamefont
			{Noh}}, \bibinfo {author} {\bibfnamefont {M.~H.}\ \bibnamefont {Upton}},
		\bibinfo {author} {\bibfnamefont {D.}~\bibnamefont {Casa}}, \bibinfo {author}
		{\bibfnamefont {T.}~\bibnamefont {Gog}}, \bibinfo {author} {\bibfnamefont
			{A.}~\bibnamefont {Paramekanti}}, \ and\ \bibinfo {author} {\bibfnamefont
			{Y.-J.}\ \bibnamefont {Kim}},\ }\href {\doibase 10.1103/PhysRevB.95.235114}
	{\bibfield  {journal} {\bibinfo  {journal} {Phys. Rev. B}\ }\textbf {\bibinfo
			{volume} {95}},\ \bibinfo {pages} {235114} (\bibinfo {year}
		{2017})}\BibitemShut {NoStop}%
	\bibitem [{\citenamefont {Kobayashi}\ \emph {et~al.}(2011)\citenamefont
		{Kobayashi}, \citenamefont {Nagao},\ and\ \citenamefont
		{Ito}}]{Kobayashi2011}%
	\BibitemOpen
	\bibfield  {author} {\bibinfo {author} {\bibfnamefont {K.}~\bibnamefont
			{Kobayashi}}, \bibinfo {author} {\bibfnamefont {T.}~\bibnamefont {Nagao}}, \
		and\ \bibinfo {author} {\bibfnamefont {M.}~\bibnamefont {Ito}},\ }\href
	{\doibase 10.1107/S010876731102633X} {\bibfield  {journal} {\bibinfo
			{journal} {Acta Crystallographica Section A}\ }\textbf {\bibinfo {volume}
			{67}},\ \bibinfo {pages} {473} (\bibinfo {year} {2011})}\BibitemShut
	{NoStop}%
	\bibitem [{\citenamefont {Stassis}\ and\ \citenamefont
		{Deckman}(1976)}]{Stassis1976}%
	\BibitemOpen
	\bibfield  {author} {\bibinfo {author} {\bibfnamefont {C.}~\bibnamefont
			{Stassis}}\ and\ \bibinfo {author} {\bibfnamefont {H.~W.}\ \bibnamefont
			{Deckman}},\ }\href {\doibase 10.1088/0022-3719/9/12/008} {\bibfield
		{journal} {\bibinfo  {journal} {Journal of Physics C: Solid State Physics}\
		}\textbf {\bibinfo {volume} {9}},\ \bibinfo {pages} {2241} (\bibinfo {year}
		{1976})}\BibitemShut {NoStop}%
	\bibitem [{\citenamefont {Pourovskii}\ \emph {et~al.}(2020)\citenamefont
		{Pourovskii}, \citenamefont {Boust}, \citenamefont {Ballou}, \citenamefont
		{Eslava},\ and\ \citenamefont {Givord}}]{Pourovskii2020}%
	\BibitemOpen
	\bibfield  {author} {\bibinfo {author} {\bibfnamefont {L.~V.}\ \bibnamefont
			{Pourovskii}}, \bibinfo {author} {\bibfnamefont {J.}~\bibnamefont {Boust}},
		\bibinfo {author} {\bibfnamefont {R.}~\bibnamefont {Ballou}}, \bibinfo
		{author} {\bibfnamefont {G.~G.}\ \bibnamefont {Eslava}}, \ and\ \bibinfo
		{author} {\bibfnamefont {D.}~\bibnamefont {Givord}},\ }\href {\doibase
		10.1103/PhysRevB.101.214433} {\bibfield  {journal} {\bibinfo  {journal}
			{Phys. Rev. B}\ }\textbf {\bibinfo {volume} {101}},\ \bibinfo {pages}
		{214433} (\bibinfo {year} {2020})}\BibitemShut {NoStop}%
	\bibitem [{\citenamefont {Amadon}\ \emph {et~al.}(2008)\citenamefont {Amadon},
		\citenamefont {Lechermann}, \citenamefont {Georges}, \citenamefont {Jollet},
		\citenamefont {Wehling},\ and\ \citenamefont {Lichtenstein}}]{Amadon2008}%
	\BibitemOpen
	\bibfield  {author} {\bibinfo {author} {\bibfnamefont {B.}~\bibnamefont
			{Amadon}}, \bibinfo {author} {\bibfnamefont {F.}~\bibnamefont {Lechermann}},
		\bibinfo {author} {\bibfnamefont {A.}~\bibnamefont {Georges}}, \bibinfo
		{author} {\bibfnamefont {F.}~\bibnamefont {Jollet}}, \bibinfo {author}
		{\bibfnamefont {T.~O.}\ \bibnamefont {Wehling}}, \ and\ \bibinfo {author}
		{\bibfnamefont {A.~I.}\ \bibnamefont {Lichtenstein}},\ }\href@noop {}
	{\bibfield  {journal} {\bibinfo  {journal} {Phys. Rev. B}\ }\textbf {\bibinfo
			{volume} {77}},\ \bibinfo {pages} {205112} (\bibinfo {year}
		{2008})}\BibitemShut {NoStop}%
\end{thebibliography}

%


\end{document}


\title{Supplementary material for 
'Ferro-octupolar order and low-energy excitations in d$^2$ double perovskites of Osmium'}

\author{Leonid V. Pourovskii}
\address{Centre de Physique Th\'eorique, Ecole Polytechnique, CNRS, Institut Polytechnique de Paris,
  91128 Palaiseau Cedex, France}
\address{Coll\`ege de France, 11 place Marcelin Berthelot, 75005 Paris, France}

\author{Dario Fiore Mosca}
\address{University of Vienna, Faculty of Physics and Center for Computational Materials Science, Vienna, Austria}

\author{Cesare Franchini}
\address{University of Vienna, Faculty of Physics and Center for Computational Materials Science, Vienna, Austria}
\address{Department of Physics and Astronomy, Alma Mater Studiorum - Universit\`a di Bologna, Bologna, 40127 Italy}

\date{\today}	
\maketitle

\section{First principles methods}

\subsection{DFT+HI}
In order to evaluate the effective Hamiltonian from first principles, we start by calculating the electronic structure of  paramagnetic Ba$_2M$OsO$_6$ with the DFT+dynamical mean-field theory(DMFT) method. {  The quantum impurity problem for the whole Os 5$d$ shell} is solved within the quasi-atomic Hubbard-I (HI) approximation \cite{hubbard_1}; the method is abbreviated below as DFT+HI.  We employ a self-consistent DFT+DMFT implementation\cite{Aichhorn2009,Aichhorn2011,Aichhorn2016} based on the full-potential LAPW code Wien2k\cite{Wien2k} and including the spin-orbit with the standard variational treatment. Wannier orbitals representing Os 5$d$ orbitals are constructed from the Kohn-Sham (KS) bands in the energy range [-1.2:6.1]~eV relative to the KS Fermi level; this energy window includes all $t_{2g}$ and $e_g$ states of Os but not the oxygen 2$p$ bands. 

\lp{The whole Os 5$d$ shell is retained in the DMFT impurity problem, with the on-site Coulomb repulsion defined for this shell using the Slater parameters $F^0$, $F^2$ and $F^4$ together with a standard assumption\cite{Anisimov1993} for the ratio $F^4/F^2$=0.625.  The on-site Coulomb vertex in this case is fully determined by two parameters: $U=F^0$ and $J_H=(F^2+F^4)/14$.} 
We use $U=F_0=3.2$~eV for the BMOO and BZOO; for BCOO we employ a slightly larger value of $U=$3.5 eV to stabilize the $d^2$ ground state in DFT+HI. We use   $J_H$=0.5~eV for all three compounds.  Our values for $U$ and $J_H$ are consistent with previous calculations of d$^1$ Os perovskites by DFT+HI\cite{FioreMosca2021} \lp{and with experimental estimates (e.~g.  U$\sim$3.2~eV by Ref.~\cite{Erickson2007}). The Kanamori parametrization  is often employed in the literature to define the $t_{2g}$ Coulomb vertex; the corresponding Kanamori Hund's rule coupling $J_H^K\approx0.77 J_H$\cite{Georges_Hunds_materials}. Our value of $J_H=$0.5~eV thus corresponds to $J_H^K=$0.385~eV; it is within the range of experimental estimates of $J_H^K=0.2$ to 0.4~eV for 5d ions\cite{Winter2016,Taylor2017,Yan2017}.  We have also performed test calculations for BZOO with $J_H$ varied in the range from 0.2 to 0.5~eV; changing $J_H$ does not affect the qualitative picture of a ferro-octupolar ground state.}

All calculations are carried out for the experimental cubic lattice structures of Ba$_2M$OsO$_6$, the lattice parameter $a=$8.346, 8.055, and 8.082~\AA\ for $M=$Ca, Mg, and Zn, respectively\cite{Thompson2014,Marjerrison2016}. We employ the local density approximation as the DFT exchange-correlation potential, 400 $\vk$-point in the full Brillouin zone, and the Wien2k basis cutoff $R_{\mathrm{mt}}K_{\mathrm{max}}=$8. The double-counting correction is evaluated using the fully-localized limit with the nominal 5$d$ shell occupancy of 2.

\subsection{Calculations of inter-site exchange interactions (IEI)}
In order to evaluate all IEI $V_{KK'}^{QQ'}(\Delta\vR)$ acting within $J_{eff}$=2 manifold,  we employ the HI-based force-theorem approach of Ref.~\onlinecite{Pourovskii2016} (abbreviated below as FT-HI). Within this approach, the matrix elements of IEI   $V(\Delta\vR)$ coupling $J_{eff}$=2 shells on two Os sites  read
\beq\label{V}
\langle M_1 M_3| V(\Delta\vR)| M_2 M_4\rangle=\mathrm{Tr} \left[ G_{\Delta\vR}\frac{\delta\Sigma^{at}_{\vR+\Delta\vR}}{\delta \rho^{M_3M_4}_{\vR+\Delta\vR}} G_{-\Delta\vR}\frac{\delta\Sigma^{at}_{\vR}}{\delta \rho^{M_1M_2}_{\vR}}\right],
\eeq
where  $\Delta\vR$ is the lattice vector connecting the two sites,  $M=-2...2$ is the projection quantum number,  $\rho^{M_iM_j}_{\vR}$ is the corresponding element of the $J_{eff}$ density matrix on site $\vR$, $\frac{\delta\Sigma^{at}_{\vR}}{\delta \rho^{M_iM_j}_{\vR}}$ is the derivative of atomic (Hubbard-I) self-energy $\Sigma^{at}_{\bf R}$ over a fluctuation of the $\rho^{M_iM_j}_{\vR}$ element, $G_{\vR}$ is the inter-site Green's function.  
The self-energy derivatives are calculated with analytical formulas from atomic Green's functions.
The FT-HI method is applied as a post-processing on top of DFT+HI, hence, all quantities in the RHS of eq.~\ref{V} are evaluated from the fully converged DFT+HI electronic structure of a given system.

Once all matrix elements (\ref{V}) are calculated, they are directly mapped into the corresponding couplings $V_{KK'}^{QQ'}(\Delta\vR)$ between on-site moments (eq. 22 in Ref.~\onlinecite{Pourovskii2016}). To have a correct mapping into the $J_{eff}$ pseudo-spin basis the phases of $|J_{eff}M\rangle$ are aligned, i.~e. they are chosen such that $\langle J_{eff}M|J_{+}|J_{eff}M-1\rangle$ is a  real positive number. 

The calculations of IEI within the $E_g$ space proceed in the same way starting from the same converged DFT+HI electronic structure. The density matrices fluctuations $\rho^{M_iM_j}_{\vR}$ are  restricted within  the $E_g$ doublet, and $M=\pm \frac{1}{2}$. 

In the converged DFT+HI electronic structure the chemical potential $\mu$ is sometimes found to be pinned at the very top of the valence (lower Hubbard) band instead of being strictly inside the Mott gap. Since the FT-HI method breaks down if any small metallic spectral weight  is present, in those cases we calculated the IEI with $\mu$ shifted into the gap.

\subsection{Generalized dynamical susceptibility.} 
We evaluated the generalized dynamical susceptibility in the FO $xyz$ ordered state using the random phase approximation (RPA), see, e.~g., Ref.~\cite{RareEarthMag_book}.  
Within the RPA, the general susceptibility matrix in the $J_{eff}$=2 space reads
  \beq\label{eq:chi}
  \bar{\chi}(\vq,E)=\left[I-\bar{\chi}_0(E)\bar{V}_{\vq}\right]^{-1}\bar{\chi}_0(E),
  \eeq
  where $\bar{\chi}_0(E)$ is the on-site bare susceptibility, $\bar{V}_{\vq}$ is the Fourier transform of IEI matrices $\hat{V}(\Delta \vR)$, the bar $\bar{...}$ designates a matrix in the combined $\mu=[K,Q]$ index labeling $J_{eff}$ multipoles Notice, that $\hat{V}(\Delta \vR)$ and, correspondingly, $V_{\vq}$ do not couple time-odd and time-even multipoles. The on-site susceptibility $\hat{\chi}_{0}(E)$  is calculated  in accordance with eq.~3 of the main text.
  
%

  	\renewcommand\floatpagefraction{0.1}
  \begin{table}[tp]
  	\caption{\label{Tab:SEI}  
  		 Calculated IEI $V_{KK'}^{QQ'}$ for the $J_{eff}$=2  multiplet. First two columns list
  		$Q$ and $Q'$ , respectively. Third and fourth column displays the $KQ$ and $K'Q'$ tensors in the  Cartesian representation, respectively. The three last columns display the values of IEI for BCOO, BMOO, and BZOO in meV.
  	}
	\begin{center}
		\begin{ruledtabular}
			\renewcommand{\arraystretch}{1.2}
			\begin{tabular}{c c c c c c c }
  & & & {Dipole-Dipole} & BCOO & BMOO & BZOO \\
 				\hline
-1 & -1   & y  & y  &    1.62  &    1.47  &    0.97  \\
0 &  0   & z  & z  &    4.17  &    4.12  &    2.96  \\
1 & -1   & x  & y  &    1.26  &    1.42  &    1.02  \\
1 &  1   & x  & x  &    1.62  &    1.47  &    0.97  \\
		\hline
		\hline
\multicolumn{7}{c}{Quadrupole-Quadrupole} \\
\hline
-2 & -2   & xy  & xy  &   -0.41  &   -0.52  &   -0.31  \\
-1 & -1   & yz  & yz  &   -0.79  &   -0.75  &   -0.60  \\
0 & -2   & z$^2$  & xy  &    0.16  &    0.07  &    0.07  \\
0 &  0   & z$^2$  & z$^2$  &    1.32  &    1.49  &    0.99  \\
1 & -1   & xz  & yz  &   -0.23  &   -0.22  &   -0.19  \\
1 &  1   & xz  & xz  &   -0.79  &   -0.75  &   -0.60  \\
2 &  2   & x$^2$-y$^2$  & x$^2$-y$^2$  &   -0.58  &   -0.65  &   -0.48  \\
\hline
		\hline
\multicolumn{7}{c}{Octupole-Octupole} \\
\hline
-3 & -3   & y(3x$^2$-y$^2$)  & y(3x$^2$-y$^2$)  &    1.16  &    1.25  &    1.24  \\
-2 & -2   & xyz  & xyz  &   -1.49  &   -1.47  &   -0.85  \\
-1 & -3   & yz$^2$  & y(3x$^2$-y$^2$)  &   -0.14  &   -0.19  &   -0.21  \\
-1 & -1   & yz$^2$  & yz$^2$  &    0.80  &    0.81  &    0.33  \\
0 & -2   & z$^3$  & xyz  &   -0.79  &   -0.88  &   -0.57  \\
0 &  0   & z$^3$  & z$^3$  &    2.35  &    2.42  &    1.33  \\
1 & -3   & xz$^2$  & y(3x$^2$-y$^2$)  &   -0.29  &   -0.37  &   -0.38  \\
1 & -1   & xz$^2$  & yz$^2$  &   -0.98  &   -1.12  &   -0.82  \\
1 &  1   & xz$^2$  & xz$^2$  &    0.80  &    0.81  &    0.33  \\
2 &  2   & z(x$^2$-y$^2$)  & z(x$^2$-y$^2$)  &   -1.89  &   -2.00  &   -1.42  \\
3 & -1   & x(x$^2$-3y$^2$)  & yz$^2$  &    0.29  &    0.37  &    0.38  \\
3 &  1   & x(x$^2$-3y$^2$)  & xz$^2$  &    0.14  &    0.19  &    0.21  \\
3 &  3   & x(x$^2$-3y$^2$)  & x(x$^2$-3y$^2$)  &    1.16  &    1.25  &    1.24  \\
		\hline
		\hline
\multicolumn{7}{c}{Dipole-Octupole} \\
\hline
-1 & -3   & y  & y(3x$^2$-y$^2$)  &     &   -0.07  &   -0.06  \\
-1 & -1   & y  & yz$^2$  &    1.97  &    1.92  &    0.93  \\
-1 &  1   & y  & xz$^2$  &   -0.20  &   -0.23  &   -0.27  \\
-1 &  3   & y  & x(x$^2$-3y$^2$)  &   -0.89  &   -1.01  &   -0.99  \\
0 & -2   & z  & xyz  &   -0.97  &   -1.29  &   -1.19  \\
0 &  0   & z  & z$^3$  &    2.38  &    2.19  &    1.04  \\
1 & -3   & x  & y(3x$^2$-y$^2$)  &    0.89  &    1.01  &    0.99  \\
1 & -1   & x  & yz$^2$  &   -0.20  &   -0.23  &   -0.27  \\
1 &  1   & x  & xz$^2$  &    1.97  &    1.92  &    0.93  \\
1 &  3   & x  & x(x$^2$-3y$^2$)  &     &    0.07  &    0.06  \\
  			\end{tabular}
\end{ruledtabular}
\end{center}
\end{table}

\section{Intersite exchange interactions in the $J_{eff}$=2 space}\label{sec:IEI_table}
The IEI between $J_{eff}$=2 multipoles for a pair of Os sites form a 24$\times$24 matrix  $\hat{V}(\Delta\vR)$, since  $K_{max}(K_{max}+2)$=24 with $K_{max}=2J_{eff}$. 
In Supplementary Table~\ref{Tab:SEI} we list all calculated IEI  in the three systems with magnitude above 0.05 meV. The IEI are given for the [0.5,0.5,0.0] Os-Os nearest-neighbor lattice vector.  The calculated next-nearest-neighbor interactions  are at least one order of magnitude smaller than the NN one; longer range IEI were neglected. 

The IEI between hexadecapoles as well as between hexadecapoles and quadrupoles are below this cutoff and not shown. The same applies to the next-nearest-neighbour IEI, which are all below 0.05 meV in the absolute value.

\section{Projection of $J_{eff}$=2 multipolar operators into the $E_g$ space}
In the $|J_{eff}=2,M\rangle$  basis the pseudo-spin-1/2 states of $E_g$ ground-state doublet read
\beq\label{eq:Eg_states}
|\uparrow\rangle=|2,0\rangle;\\ 
|\downarrow\rangle=\left(|2,-2\rangle+|2,2\rangle\right)/\sqrt{2}.
\eeq
 The resulting $E_g$ Hamiltonian is then related to the $J_{eff}$=2 one (eq. 1 of main text) by the projection 
\beq\label{eq:H_Eg}
H_{E_g}=\hat{P}H_{IEI}\hat{P}^T=\sum_{\langle ij\rangle \in NN}\sum_{\alpha\beta}J_{\alpha\beta}(\Delta \vR_{ij})\tau_{\alpha}(\vR_i)\tau_{\beta}(\vR_j),
\eeq
 where the rows of projection matrix $P$ are the $E_g$ states in $J_{eff}=2$ basis (\ref{eq:Eg_states}), $\tau_{\alpha}$ is the spin-1/2 operator for $\alpha=x$,$y$, or $z$.

Only six $J_{eff}$=2 multipoles out of 24 have non-zero projection into the $E_g$ space; those projections expanded into the spin-1/2 operators are listed below. Namely, there are two quadrupoles 
$$
O_2^0 \equiv O_{z^2} \to 2\sqrt{2/7} \tau_z, \; \; \;
O_2^2\equiv O_{x^2-y^2}\to 2\sqrt{2/7}\tau_x, 
$$
the $xyz$ octupole
$$
O_3^{-2}\equiv O_{xyz}\to -\sqrt{2} \tau_y, 
$$ 
as well as three hexadecapoles 
$$
O_4^0\to \sqrt{7/40}I-\sqrt{5/14}\tau_z, \; \; \;
O_4^2\to \sqrt{6/7} \tau_x, \; \; \;
O_4^4\to (1/\sqrt{8})I+(1/\sqrt{2})\tau_z.
$$
The $O_4^0$ and $O_4^4$ hexadecapoles contribute to the remnant CF $H_{rcf}$; they have, correspondingly,  non-zero traces in the $E_g$ space. Hence the presence of "monopole" (unit $2 \times 2$ matrix) $I$ in their projections to $E_g$.
To simplify subsequent expressions one may transform the hexadecapolar operators into a symmetry-adapted basis:
\beq\label{eq:trans_O4}
\begin{pmatrix}
	O^I_4 \\ O^z_4
\end{pmatrix}
=
\begin{pmatrix}
	\cos \theta & \sin \theta  \\  -\sin \theta & \cos \theta   
\end{pmatrix}
\begin{pmatrix}
	O^0_4 \\ O^4_4
\end{pmatrix},
\eeq
where $\theta=\arccos(\sqrt{7/12})$. The transformed operators have the following projections into the $E_g$ space:
$$
O_4^I\to \sqrt{3/10}I, \; \; \;
O^z_4 \to \sqrt{6/7}\tau_z .
$$

Substituting those expressions for the relevant multipoles into the effective Hamiltonian  $H_{eff}$ (eq.~1 of the main text) one may derive explicit formulas for the $E_g$ IEI in terms of the $J_{eff}$=2 IEI:
\begin{gather}
J_{yy}=2 V_{33}^{\bar{2}\bar{2}}, \\
J_{zz}=2\left[\frac{4}{7}V_{20}^{20}+\frac{4\sqrt{3}}{7}V_{24}^{2z}+\frac{3}{7}V_{44}^{zz}\right], \\
J_{xx}=2\left[\frac{4}{7}V_{22}^{22}+\frac{4\sqrt{3}}{7}V_{24}^{22}+\frac{3}{7}V_{44}^{22}\right],
\end{gather}
where we drop the $\vR$ argument in $V_{KK'}^{QQ'}(\vR)$ for brevity. $V_{24}^{2z}$ and $V_{44}^{zz}$ are the IEI transformed to the symmetry-adapted basis (\ref{eq:trans_O4}). The overall prefactor 2 is due to   different normalizations of the spin operators and the  spherical tensors. 

One sees that the $xyz$ octupole IEI directly maps into $J_{yy}$. In contrast, $J_{zz}$ and $J_{xx}$ are combinations of quadrupole and hexadecapole IEI. Since the IEI involving hexadecapoles are small (see Sec.~\ref{sec:IEI_table}), $J_{xx}$ and $J_{zz}$  are essentially given by the two corresponding quadrupolar IEI in the $J_{eff}$=2 space. 
However, the admixture of hexadecapole IEI into $J_{xx}$ and $J_{yy}$ leads to a reduced prefactor for the quadrupole contributions.
Hence, one sees that $J_{yy}$ is equal to 2$V_{33}^{\bar{2}\bar{2}}$, while $J_{xx}$ and $J_{zz}$  are essentially given by 8/7 of the corresponding quadrupolar couplings, $V_{22}^{22}$ and $V_{22}^{00}$,  in $J_
{eff}$=2. 

By comparing the data in Table~I of the main text  with Supp. Table~\ref{Tab:SEI} one see that this result holds for the IEI evaluated numerically using the FT-HI approach.

\section{Formalism for the inelastic neutron-scattering (INS) cross-section beyond the dipole approximation}

\subsection{INS cross-section through generalized multipolar susceptibility}

We start with the general formula for the magnetic neutron-scattering cross-section\cite{Lovesey_book_full,RareEarthMag_book} from a lattice of atoms:
\beq\label{eq:Xsec_gen}
\frac{d^2 \sigma}{d \Omega dE'}=r_0^2\frac{k'}{k}\sum_{n,n'}P_n|\langle n'|\hat{\vQ}_t^{\perp}(\vq)|n\rangle|^2\delta(\hbar\omega+E_n-E_{n'}),
\eeq
where  $r_0=$-5.39$\cdot$10$^{-13}$~cm is the characteristic magnetic scattering length,  $k$ and $k'$ are the magnitudes of initial and final neutron momentum, $|n\rangle$ and $|n'\rangle$ are the initial and final electronic states of the lattice, $E_n$ and $E_{n'}$ are the corresponding energies, $P_n$ is the probability for the lattice to be in the initial state $|n\rangle$.  We consider the case of INS with the energy transfer to the system $\hbar\omega \ne $0. Finally,  $\hat{\vQ}_t^{\perp}(\vq)$ is the neutron scattering operator, which is a sum of single-site contributions:
\beq\label{eq:Qlat}
\hat{\vQ}_t^{\perp}(\vq)=\vq\times \left(\sum_i \hat{\vQ}_i(\vq) e^{i\vq\vR_i}\right) \times \vq.
\eeq
The single-site one-electron operator $\vQ_i(\vq)$ at the site $i$ reads
\beq\label{eq:Qi}
\hat{\vQ}_i(\vq)=\hat{\vQ}_{is}(\vq)+\hat{\vQ}_{io}(\vq)=\sum_j e^{i\vq \vr_j}\left[\hat{\vs}_j-\frac{i}{q^2}(\vq\times \hat{\vp}_j)\right],
\eeq
where the sum includes all electrons on a partially-filled shell, $\vr_j$ is the position of electron $j$ on this shell with respect to  the position $\vR_i$ of this lattice site, $\hat{\vp}_j$ is the momentum operator acting on this electron. The on-site operator $\hat{\vQ}_i$  consists of the spin $\hat{\vQ}_{is}(\vq)$ and orbital $\hat{\vQ}_{io}(\vq)$ terms. We  note that $\hat{\vQ}$, as any one-electron operator acting within an atomic multiplet with the total momentum $J$, can be decomposed into many-electron multipole operators of that multiplet\cite{Shiina2007}:
\beq\label{eq:Q_in_multipoles}
\hat{Q}_i^{\alpha}(\vq)=\sum_{\mu}F_{\alpha\mu}(\vq)O_{\mu}(\vR_i),
\eeq
where  $\alpha=x$, $y$, or $z$, $O_{\mu}(\vR_i)$ is the  spherical tensor operator for the multipole $\mu\equiv \{K,Q\}$ of the total momentum $J$ acting on the site $i$, and $F_{\alpha\mu}(\vq)$ is the corresponding form-factor. In contrast to the usual dipole form-factors depending only on the magnitude $q$ of the momentum transfer, for a general multipole $\mu$ it may also depend on the momentum transfer's direction. Since $\hat{\vQ}$ is a time-odd operator, only multipoles with odd $K$ contribute into (\ref{eq:Q_in_multipoles}).

By inserting  (\ref{eq:Q_in_multipoles}) into (\ref{eq:Qlat}) and then the resulting expression for $\hat{\vQ}_t^{\perp}(\vq)$ into (\ref{eq:Xsec_gen}), one obtains an expression for the magnetic INS cross-section through the form-factors and matrix elements of the multipole operators:
\beq\label{eq:Xsec_multipoles}
\frac{d^2 \sigma}{d \Omega dE'}=r_0^2\frac{k'}{k}\sum_{ii'}\sum_{n,n'}P_n
\langle n|\vq\times\left(\sum_{\mu}\vF_{\mu}O_{\mu}(\vR_i)\right)\times\vq|n'\rangle
\langle n'|\vq\times\left(\sum_{\mu'}\vF_{\mu'}O_{\mu'}(\vR_{i'})\right)\times\vq|n\rangle
\delta(\hbar\omega+E_n-E_{n'}),
\eeq
where $\vF_{\mu}$ is the 3D vector of form-factors for the multipole $\mu$. Finally,  using the same steps as in the standard derivation of the cross-section  within the dipole approximation \cite{RareEarthMag_book} we obtain the following expression for the magnetic INS cross-section of non-polarized neutrons:
\beq\label{eq:Xsec_multipoles}
\frac{d^2 \sigma}{d \Omega dE'}=r_0^2\frac{k'}{k}\sum_{\alpha\beta}\left(\delta_{\alpha\beta}-q_{\alpha}q_{\beta}\right) \\ \left[\sum_{\mu\mu'}F_{\alpha\mu}(\vq)F_{\beta\mu'}(\vq) \frac{1}{2\pi\hbar}S_{\mu\mu'}(\vq,E)\right],
\eeq
where the dynamic correlation function  $S_{\mu\mu'}(\vq,E)$  for $\vq$ and the energy transfer $E=\hbar\omega$ is related to the generalized susceptibility $\bar{\chi}(\vq,E)$ (eq.~\ref{eq:chi} above) by the fluctuation-dissipation theorem:
\beq\label{eq:FDT}
S_{\mu\mu'}(\vq,E)=\frac{2\hbar}{1-e^{-E/T}}\chi''_{\mu\mu'}(\vq,E),
\eeq
where $T$ is the temperature, and the absorptive part of susceptibility $\chi''_{\mu\mu'}(\vq,E)=\mathrm{Im}\chi_{\mu\mu'}(\vq,E)$ in the relevant case of a cubic lattice structure with the inversion symmetry. We then insert  (\ref{eq:FDT}) into (\ref{eq:Xsec_multipoles}) omitting the detalied-balance prefactor $1/(1-e^{-E/T})\approx 1$ for the present case of a near-zero temperature and a large excitation gap. We also omit the constant prefactors and the ratio $k'/k$, which depends on the initial neutron energy in experiment, and thus obtain eq. 5 of the main text.

\subsection{Calculations of the form-factors}

In order to evaluate the form-factors $F_{\alpha\mu}(\vq)$ one needs the matrix elements 
\beq\label{eq:1el_mel}
\langle lms|\hat{\vQ}(\vq)|lm's'\rangle 
\eeq
of the one-electron neutron scattering operator (\ref{eq:Qi}) for the 5$d$ shell ($l$=2) of Os$^{6+}$ ($l$, $m$, and $s$ are the orbital, magnetic and spin quantum numbers of one-electron orbitals, respectively). We compute those matrix elements employing the analytical expressions for the spin and orbital part of $\hat{\vQ}(\vq)$ that are derived in chap. 11 of the book by Lovesey \cite{Lovesey_book_full}; they are  succinctly summarized  by Shiina {\it et al.}~\cite{Shiina2007}. Notice that in eqs. 13 and 14 of Ref.~\onlinecite{Shiina2007}  the matrix elements are given for the  projected operator $\vq\times\hat{\vQ}(\vq)\times\vq$, but they are quite simply related to those of unprojected $\hat{\vQ}(\vq)$ (see also eq.~11.48 in Ref.~\onlinecite{Lovesey_book_full}).  The radial integrals $\langle j_L(q)\rangle$ for the Os$^{6+}$ 5$d$ shell, which enter into the formulas for one-electron matrix elements,  were taken from Ref.~\onlinecite{Kobayashi2011} .  

  \begin{figure}[!tb]
	\begin{centering}
		\includegraphics[width=0.85\columnwidth]{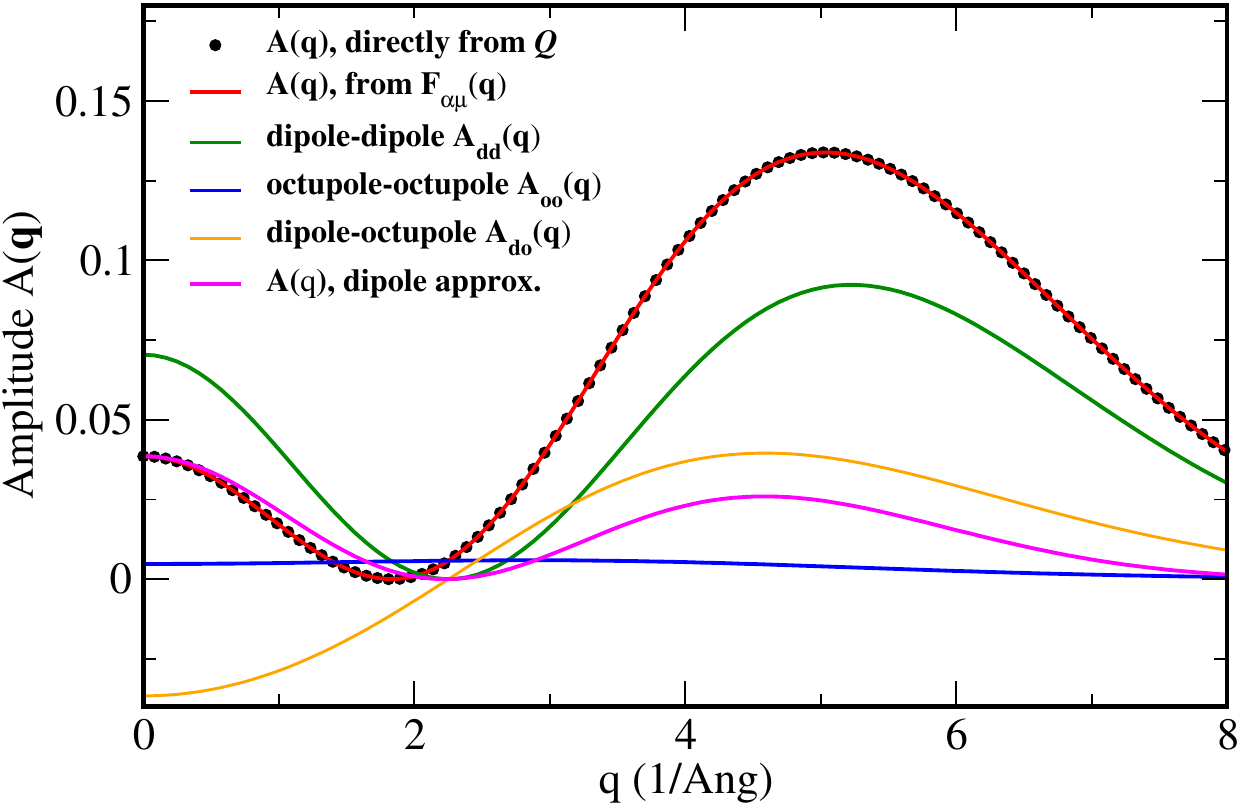} 
		\par\end{centering}
	\caption{$\vq$-dependent prefactor (\ref{eq:q_pref}) for the elastic neutron scattering along the direction [100] in the $\vq$ space, for the saturated $|22\rangle$ state of the Os$^{6+}$ $J_{eff}$=2 multiplet. The meaning of various curves is explained in the text}
	\label{fig:FF_JJstate} 
\end{figure}

In order to obtain matrix elements of $\hat{\vQ}(\vq)$  for many-electron states from the one-electron ones (\ref{eq:1el_mel}), Refs.~\onlinecite{Stassis1976,Lovesey_book_full,Shiina2007} generally assume a certain coupling scheme for a given ion ($LS$ or $jj$). Instead we simply use the atomic two-electron states of Os$^{6+}$ $J_{eff}$=2 shell as obtained by converged DFT+HI for  a given Ba$_2M$OsO$_6$ system to calculate those matrix elements numerically for each point of the $\vq$-grid. Since the two-electron atomic eigenstates are expanded in the Fock space of $(lms)$ orbitals, such calculation is trivial. The resulting matrices in the $J_{eff}$ space with matrix elements  $Q_{\alpha}^{MM'}(\vq)=\langle J_{eff}M| \hat{Q}_{\alpha}(\vq) |J_{eff}M'\rangle$ are then expanded in the odd $J_{eff}$ multipoles in accordance with (\ref{eq:Q_in_multipoles}) to obtain the form-factors $F_{\alpha\mu}(\vq)$ for each direction $\alpha$.

\subsection{Form-factors for the saturated $M=J$ state of the $J_{eff}$=2 multiplet}

As an example of application of the approach described above, let us consider the neutron-scattering form-factors for  the saturated $|J=2,M=J\rangle\equiv |JJ\rangle$ state of the  Os$^{6+}$  5d$^2$ $J_{eff}$=2 multiplet. We evaluate the corresponding $\vq$-dependent prefactor for elastic scattering
\beq\label{eq:q_pref}
A(\vq)=\sum_{\alpha\beta}\left(\delta_{\alpha\beta}-q_{\alpha}q_{\beta}\right) \langle \hat{Q}_{\alpha}(\vq)\rangle_{JJ} \langle \hat{Q}_{\beta}(\vq)\rangle_{JJ} ,
\eeq
 for the case of $|JJ\rangle$ ground state (which is, of course, not  realized in the actual Ba$_2M$OsO$_6$ systems); $\hat{Q}_{\alpha}(\vq)$ is the neutron-scattering operator (\ref{eq:Qi}) for the direction $\alpha$, by $\langle \hat{X}\rangle_{JJ}$ we designate the expectation value of an operator $X$ in  the $|JJ\rangle$ state, $\langle \hat{X}\rangle_{JJ}\equiv\langle JJ|\hat{X}|JJ\rangle$ .  We consider $\vq $ along the [100] direction; corresponding $A(\vq)$ vs. $q$  obtained by direct evaluation of the matrix elements using (\ref{eq:1el_mel}) is shown in Supplementary  Fig.~\ref{fig:FF_JJstate} by dots.  It can be compared with the same prefactor (shown in Supplementary  Fig.~\ref{fig:FF_JJstate}  in magenta) calculated within the dipole approximation for matrix elements:
 \beq\label{eq:dip_app}
 \langle \hat{\vQ}(\vq)\rangle_{JJ} \simeq \frac{1}{2}\left[  \langle j_0(q) \rangle \langle \mathbf{L}+2\mathbf{S}\rangle_{JJ} +  \langle j_2(q) \rangle \langle \mathbf{L} \rangle_{JJ}\right], 
 \eeq
where $L$ and $S$ are orbital and spin moment operators, respectively,  $\langle j_l(q) \rangle$ are the radial integrals\cite{Kobayashi2011} of the spherical Bessel function of order $l$ for Os$^{6+}$. Of course, within the dipole approximation (\ref{eq:dip_app}) the matrix elements of $\hat{\vQ}$ depend only on the absolute value $q$ of momentum transfer. The total $M_{tot}=\langle  \mathbf{L}+2\mathbf{S}\rangle_{JJ}$ and orbital $M_L=\langle \mathbf{L} \rangle_{JJ}$  magnetic moments  are equal to 0.39 and -1.49, respectively. The oscillatory behavior of $A(\vq)$ is thus due to $|M_{tot}| \ll |M_L|$ in  conjunction with $\langle j_0 (q) \rangle$ being  ever decreasing function and $\langle j_2(q) \rangle$ of Os$^{6+}$ peaked at non-zero $q\approx$4~\AA$^{-1}$. One may notice that $A(\vq)$ calculated beyond the dipole approximation exhibits even much stronger oscillations reaching  the overall maximum at large $q\approx$5~\AA$^{-1}$. Overall the dipole approximation is reasonable for $q <$2~\AA$^{-1}$; it underestimates very significantly the magnitude of $A(\vq)$ for larger $q$ . 

Let us now evaluate the same quantity (\ref{eq:q_pref}) using the multipole form-factors (\ref{eq:Q_in_multipoles}). The $|JJ\rangle$ state has only two non-zero odd-time multipoles: the dipole $\langle O_z \rangle_{JJ}$=0.632 and the octupole $\langle O_{z^3} \rangle_{JJ}$=0.316. For those multipoles and  $\vq||$[100] only the form-factors  for the direction $z$ are non-zero. Thus by inserting (\ref{eq:Q_in_multipoles}) into (\ref{eq:q_pref}) one obtains:
\beq\label{eq:A_in_F}
A(\vq)=F_{zz}^2(\vq)\langle O_z \rangle^2_{JJ}+F_{zz^3}^2(\vq)\langle O_{z^3} \rangle^2_{JJ}+2F_{zz}(\vq)F_{zz^3}(\vq)\langle O_z \rangle_{JJ}\langle O_{z^3} \rangle_{JJ}=A_{dd}(\vq)+A_{oo}(\vq)+A_{do}(\vq).
\eeq
One sees that the total value of $A(\vq)$ thus calculated (red line in Supplementary Fig.~\ref{fig:FF_JJstate}) coincides, as expected, with that obtained by the direct evaluation of the $\hat{Q}$ matrix elements. The advantage of using the multipole form-factors is that one may separate total $A(\vq)$  into contributions due to different multipoles and their mixtures. In the present case one obtains (Supplementary Fig.~\ref{fig:FF_JJstate}) a large oscillatory dipole contribution $A_{dd}(\vq)$, a small octupole contribution $A{oo}(\vq)$ exhibiting a shallow peak at  $q\approx$3 \AA$^{-1}$, and  mixed dipole-octupole $A_{do}(\vq)$ with the magnitude comparable to  that of $A_{dd}(\vq)$.

  \begin{figure}[!tb]
  		\includegraphics[width=0.85\columnwidth,left]{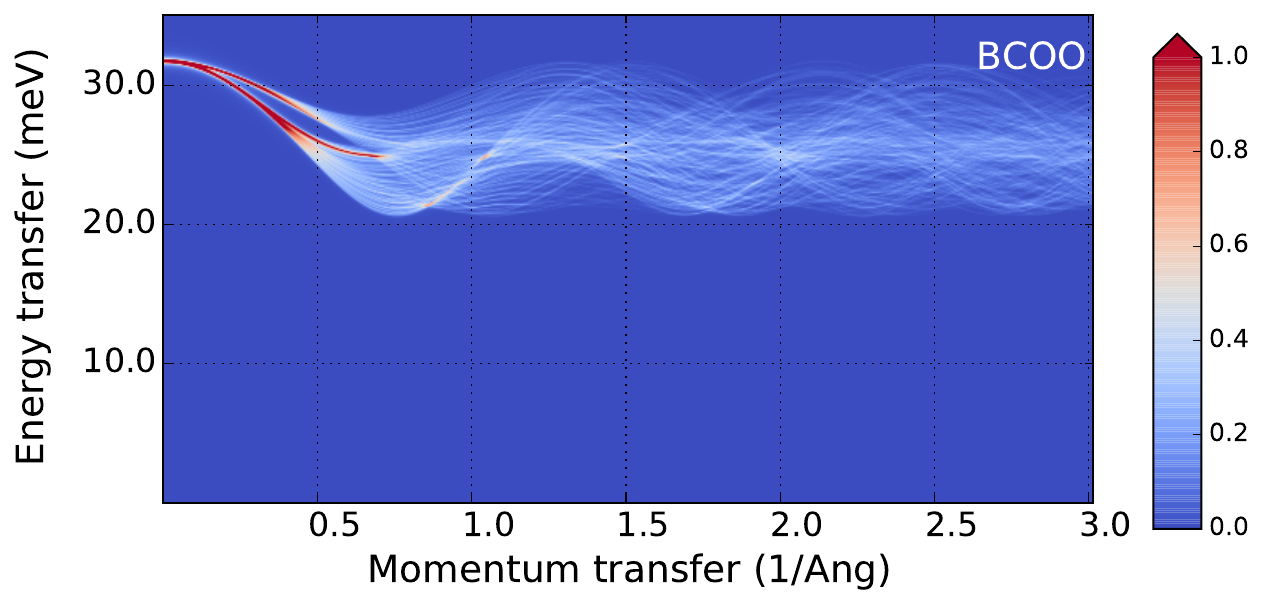}
  		\includegraphics[width=0.77\columnwidth,left]{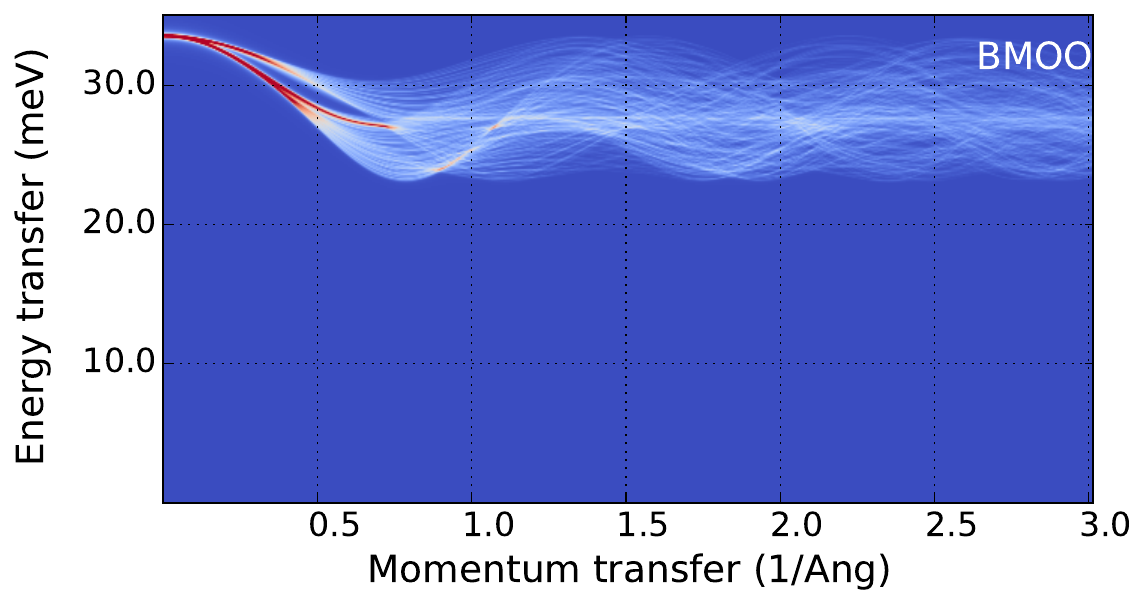}
  	\par  	    
  	\caption{Color map (in arb. units) of the calculated powder-averaged INS differential cross-section in BCOO (top) and BMOO (bottom) as a function of the energy transfer $E$ and momentum transfer $q$.}
  	\label{fig:INS_Ca_Mg} 
  \end{figure}

\section{INS cross-section of BCOO and BMOO}

In Supp. Fig.~\ref{fig:INS_Ca_Mg} we display the calculated powder-averaged INS cross-section for cubic BCOO and BMOO, the analogous data for BZOO are shown in Fig.~3a of the main text. As in the case of BZOO, only crystal-field excitations contribute to the INS, with no discernible scattering intensity present below 20 meV.

\section{Tetragonal crystal field in distorted BZOO}

In order to evaluate the dependence of tetragonal crystal field (CF) on the corresponding distortion in BZOO we have carried out self-consistent DFT+HI calculations for a set of tetragonally distorted unit cells. In these calculation we employed  the tetragonal body-centered unit cell,  which lattice parameters are $a'=a/\sqrt{2}$ and $c=a$ for an undistorted cubic lattice with the lattice parameter $a$. The tetragonal distortion was thus specified by $\delta=c/a-1=c/(\sqrt{2}a')-1$. Other parameters of those calculations ($U$, $J_H$, the choice of projection window) are the same as for the cubic structure (Supps. Sec.~I).

The local one-electron Hamiltonian for an Os 5$d$ shell in a tetragonal environment reads
\beq\label{eq:H1el}
	H_{1el}=E_0+\lambda\sum l_i s_i+L_2^0\hat{T}_2^0+L_4^0\hat{T}_4^0+L_4^4\hat{T}_4^4,
\eeq
where the first two terms in the RHS are the uniform shift and spin-orbit coupling. The last three terms represent the CF through the one-electron Hermitian Wybourne's tensors $T_k^q$ (see, e.~g., Ref.~\cite{Delange2017} for details). The term $L_2^0\hat{T}_2^0$ arises due to the tetragonal distortion. By fitting the matrix elements of (\ref{eq:H1el}) to the converged Os 5$d$ one-electron level positions as obtained by DFT+HI for a given distortion $\delta$ we extracted \cite{Delange2017} the tetragonal CF parameter $L_2^0$ vs. $\delta$. The resulting almost perfect linear dependence for small $\delta$ is shown in Supplementary Fig.~\ref{fig:L20_vs_ca}, giving $L_2^0=K'\delta$ with $K'=-13.3$~eV. 

Within the Os $d^2$ $J_{eff}$=2 multiplet the one-electron tensor $\hat{T}_2^0$ can be substituted by the corresponding Stevens operator $\hat{T}_2^0=-0.020\mathcal{O}_2^0$, where $\mathcal{O}_2^0=3J_z^2-J_{eff}(J_{eff}+1)$. In result, for the tetragonal CF parameter in the Stevens normalization $V_t=K\delta$  one obtains $K=-0.020K'=266$~meV.

  \begin{figure}[!tb]
	\begin{centering}
		\includegraphics[width=0.75\columnwidth]{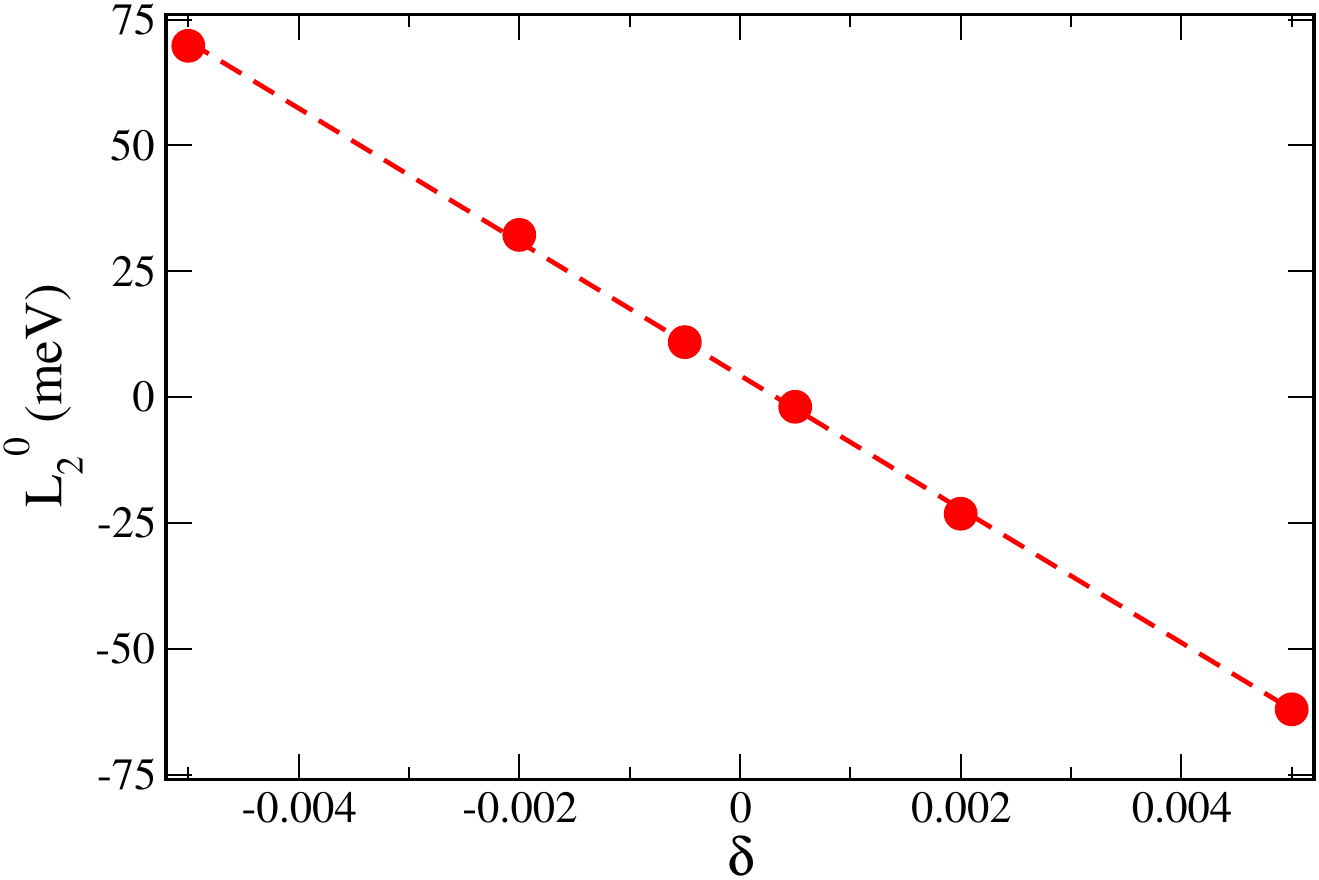} 
		\par\end{centering}
	\caption{Calculated crystal field parameter $L_{20}$ vs. tetragonal distortion $\delta=c/a-1$ in BZOO. The circles are calculated points, the line is a linear regression fit.}
	\label{fig:L20_vs_ca} 
\end{figure}

\section{Projective analysis of inter-site exchange interactions}

\subsection{Formalism}

In our DFT+HI approach the 5d states are represented by "extended" Wannier orbitals (EWO) $|w_{m\sigma}\rangle$, where $m$ and $\sigma$ are the magnetic quantum number and spin,  formed by 5d-like Os bands that  are heavily hybridized with O-2p (and, possibly, with other states).  The use of such EWO within DFT+HI allows one to effectively accounts for the hybridization (ligand) contribution to the crystal field, though this contribution is not directly included within the quasi-atomic HI approximation. Therefore, one obtains a reliable description of the crystal-field splitting in both correlated oxides \cite{Pourovskii2019,Pourovskii2021} and intermetallics\cite{Delange2017,Pourovskii2020}. However, all hopping processes between 5$d$ shells -- whether direct or indirect -- are in this case downfolded into effective hopping between those extended Os-5d states. Hence, one cannot directly separate various super-exchange and direct-exchange contributions to IEI.

In order do disentangle different contribution to inter-site exchange interactions (IEI) we thus adopted the projective approach of Ref.~\cite{Delange2017}, see Appendix F therein. Namely, we introduce a large set of  WO representing all orbitals contributing into the bands in the energy window $\mW=$[-1.2:6.1]~eV that is employed to form the EWO. We label  WO of the large set $|\tilde{w}_{\Lambda}\rangle$ introducing the combined index $\Lambda\equiv\alpha l m \sigma$, where $\alpha$ and $l$ are the atomic site and orbital quantum number, respectively. These Wannier functions in $\vk$-space are obtained from the KS bands by the projection \cite{Amadon2008,Aichhorn2009}
\beq\label{eq:LW_proj}
|\tilde{w}_{\Lambda}(\vk)\rangle=\sum_{\nu\in\tilde{\mW}} \tilde{P}_{\Lambda \nu}(\vk)|\psi_{\nu}({\vk})\rangle,
\eeq
 where $\nu$ labels KS bands  within the energy window $\tilde{\mW}$ chosen for constructing the large WO (LWO) set,  $|\psi_{\nu}({\vk})\rangle$ are the corresponding KS states. Since this set includes many orbitals it is essential to choose $\tilde{\mW}$ sufficiently wide so that it includes all corresponding bands. Within the small window $\mW$, the LWO set is approximately complete with all characters contributing to the bands within this range included. Hence, within the window $\mW$ the projection matrices $\hat{\tilde{P}}$ in (\ref{eq:LW_proj}) become essentially unitary\cite{Delange2017}, and  (\ref{eq:LW_proj}) can be inverted as
 \beq\label{eq:psi_from_w}
 |\psi_{\nu}({\vk})\rangle\approx\sum_{\Lambda} \tilde{P}^*_{\Lambda \nu}(\vk)|\tilde{w}_{\Lambda}(\vk)\rangle.
 \eeq
 In result, the two set of WO are shown to be related by a projection:
\beq\label{eq:LW_SW}
 |w_{m\sigma}(\vk)\rangle\approx\sum_{\Lambda}U_{\Lambda}^{m\sigma}(\vk)|\tilde{w}_{\Lambda}(\vk)\rangle,
 \eeq
 where the matrix elements $U_{\Lambda}^{m\sigma}(\vk)=\sum_{\nu\in\mW}P_{m\sigma\nu}(\vk)\tilde{P}^*_{\Lambda\nu}(\vk)$, $P_{m\sigma\nu}$ are the projection matrices for the EWO.
 
The projection matrix $\hat{U}$ is thus rectangular with the rows labeled by $m\sigma$ and columns labeled by $\Lambda$. We introduce 
\beq\label{eq:M_mat}
\hat{M}(\vk)=\hat{U}(\vk)\hat{U}^{\dagger}(\vk),
 \eeq
which becomes a unit matrix  in the EWO space if the condition  (\ref{eq:psi_from_w}) is fulfilled (as can be easily shown from the orthonormality of the EWO set). 

Let us now rewrite the  FT-HI eq.~(\ref{V}) for IEI through Fourier-transformed inter-site Green's functions (GF) inserting the $\hat{M}$ matrices around  each GF:
\beq\label{eq:V_in_k}
\langle M_1 M_3| V(\Delta\vR)| M_2 M_4\rangle=\sum_{\vk\vk'}e^{i(\vk'-\vk)\Delta\vR}\mathrm{Tr} \left[ M(\vk)G_{\vk}M(\vk)\frac{\delta\Sigma^{at}}{\delta \rho^{M_3M_4}} M(\vk')G_{\vk'}M(\vk')\frac{\delta\Sigma^{at}}{\delta \rho^{M_1M_2}}\right],
\eeq  
where we dropped $\vR$ labels for on-site self-energies, since in the present case all Os sites are equivalent. Defining $G^P_{\vk}=  M(\vk)G_{\vk}M(\vk)$ and performing the Fourier transforms one obtains back eq.~\ref{V} with $G_{\Delta\vR}$ substituted by 
\beq\label{eq:G_p}
G^P_{\Delta\vR}=\sum_{\vk}\hat{M}(\vk)G_{\vk}\hat{M}(\vk)e^{-i\vk \Delta\vR}.
\eeq
Obviously, if all large-window orbitals  are included and (\ref{eq:psi_from_w}) is fulfilled exactly then $G^P_{\Delta\vR}$ is simply $G_{\Delta\vR}$. The essential idea is that a subset of LWO can be excluded from the projection matrix $\hat{U}(\vk)$ by setting the corresponding columns to zero. In this case the matrix $\hat{M}$ is not a unit matrix. As one sees from eqs.~\ref{eq:LW_SW}, \ref{eq:M_mat} and   \ref{eq:G_p} this corresponds to excluding some sites/orbitals from hopping processes encoded by the inter-site GF. Hence, the resulting  IEI will be obtained with those hopping processes suppressed. In this way one may evaluate the relative importance of various contributions to the IEI.

\subsection{Results}

We constructed LWO sets for BZOO and BMOO from the Kohn-Sham band structure obtained with the converged DFT+HI charge density. We chose the orbitals to be included by inspecting the KS DOS within the window $\mW$: all characters giving a non-negligible contribution were included. In result,  Os-5d, O-2p, Ba-6s, 6p, and 5d as well as Zn-4s, 4p and 3d were included for BZOO; in the case of BMOO  Mg-3s and 3p were included instead of the Zn orbitals. We employed the windows [-9.5:12.2]~eV and [-9.5:13.6]~eV for BZOO and BMOO, respectively. We evaluated the matrices $\hat{M}(\vk)$ with all LWO included using (\ref{eq:M_mat}); the resulting matrices were almost equal to the unit one, with the diagonal elements deviating at most by about 0.01 from the unity. Hence, the LWO space is quite close to being a complete one. 

We subsequently calculated all IEI within the $J_{eff}$=2 space using those  $\hat{M}(\vk)$ in (\ref{eq:G_p}); the resulting values of the IEI relevant for the $E_g$ doublet space -- the $e_g$ quadrupole and $xyz$ octupole ones -- are listed in third column of Supp. Table~\ref{Tab:proj_IEI}.  By comparing them with the corresponding values obtained using the standard FT-HI formalism (Supp. Table~\ref{Tab:SEI}) one finds a good quantitative agreement with a deviation of at most 10\% and all qualitative tendencies well reproduced.

We then reevaluated the IEI  excluding certain orbitals from the projection matrix $\hat{U}_{\vk}$ by setting the corresponding columns to zero. This corresponds to excluding certain hopping processes from contributing into the downfolded inter-site GF $G^P_{\Delta\vR}$. The results are displayed in columns 4-7 of Supp. Table~\ref{Tab:proj_IEI}. 

In particular, by keeping only Os-5d in the LWO set one may extract the direct-exchange (DE) contribution to IEI. One sees (4th column) that the contribution due to the direct Os-5d-to-Os-5d hopping   is insignificant being less than 10\% in all cases. Hence, the  DE, which was assumed by Ref.~\cite{khaliullin2021} to be  the only important inter-site exchange in those DP, is of marginal importance. The IEI in those systems are thus due to the super-exchange (SE).

In order to estimate the relative importance of various SE processes we excluded in turn O-2p, all Ba, and all $M$-site (Zn or Mg) LWO, in each case keeping all other LWO. We note that contributions of various hopping processes into IEI are not additive (i.~e., one cannot obtain the total value by summing up contributions due to various LWO included separately). However, by excluding certain set of orbitals one may evaluate the total contribution of all SE processes involving those orbitals. Hence, to evaluate the importance of SE involving the excluded orbitals one needs to compare the IEI values calculated without them with the corresponding values obtained  including all LWO (third column). One thus finds that the SE processes involving the $M$ site (Zn or Mg) contribute little, since excluding the $M$-site orbitals modifies the IEI rather weakly. As expected, the SE involving O-2p are crucial, hence, excluding those states leads to a drastic reduction in the IEI. More interestingly, Ba orbitals seem also to provide a very significant contribution into the IEI (column 6). 

One may thus conclude that the IEI interactions in these DP are determined by SE involving oxygen O-2p and Ba states. 

Finally, we argue that the stronger IEI in BCOO and BMOO as compared to BZOO should be attributed to a more covalent Os-O bonding in the former. The reduction of Os-O covalency in BZOO is  reflected in a weaker ligand field resulting, in accordance with our DFT+HI calculations, in a 10\% smaller CF splitting between Os-5$d$ $t_{2g}$ and $e_g$ in BZOO as compared to BCOO and BMOO.  This process can be understood in the context of the so-called 'covalency competition' mechanism, namely a competition in covalent bond formation among constituent metal ions~\cite{Yamada2017}, in our case between $M$ and Os sites. Mg and Ba are chemically similar to Zn, however, Zn has a stronger tendency to form covalent bonds with the neighboring anions because of its higher electronegativity (1.65), as compared to Mg (1.31) or Ca (1.00). Consequently, the reduced degree of Os-d/O-p hybridization in BZOO weakens the principal contribution to SE, which is  due to hopping through O-2p and Ba. The corresponding enhancement of O-$M$ hybridization does not compensate for it,  since SE through $M$ site is small.

 \begin{table}[tb]
	\caption{\label{Tab:proj_IEI}  
	Calculated values of the $e_g$ quadrupole and $xyz$ octupole IEI for the [0.5,0.5,0] Os-Os nearest-neighbor lattice vector with some hopping processes excluded using the projective formalism.  The first two columns labels the interactions in the spherical tensor and Cartesian notation. The values in columns (from third to seventh) are with all LWO included, with only Os-d included, with  O-2p excluded, with all Ba orbitals excluded and with all Zn (Mg) orbitals excluded, respectively. All values are in meV.  
	}
	\begin{center}
		\begin{ruledtabular}
			\renewcommand{\arraystretch}{1.2}
			\begin{tabular}{c c c c c c c }
\multicolumn{7}{c}{{\bf BZOO}} \\
\hline
        &           & all included  &  only Os-5d   &     excl. O-2p  &    excl. Ba &      excl. Zn \\
$V_{22}^{00}$ & $V_{z^2,z^2}$     &      0.96   &      0.07   &     -0.19   &     -0.47   &      0.94   \\
$V_{22}^{22}$ & $V_{x^2-y^2,x^2-y^2}$ &     -0.50   &      0.01   &      0.03   &     -0.07   &     -0.47   \\
$V_{33}^{\bar{2}\bar{2}}$ & $V_{xyz,xyz}$   &     -0.93   &      0.04   &      0.24   &     -1.56   &     -0.86   \\
\hline
\hline
\multicolumn{7}{c}{{\bf BMOO}} \\
\hline
         &          & all included  &  only Os-5d   &     excl. O-2p  &    excl. Ba &      excl. Mg \\
$V_{22}^{00}$ & $V_{z^2,z^2}$       &      1.45   &      0.12   &     -0.29   &     -0.58   &      1.49   \\
$V_{22}^{22}$ & $V_{x^2-y^2,x^2-y^2}$ &     -0.71   &      0.03   &      0.05   &     -0.09   &     -0.77   \\
$V_{33}^{\bar{2}\bar{2}}$ & $V_{xyz,xyz}$  &     -1.68   &      0.12   &      0.33   &     -2.12   &     -1.47   \\
  			\end{tabular}
\end{ruledtabular}
\end{center}
\end{table}

\section{Anti-ferro quadrupolar order}

\begin{figure}[!tb]
	\begin{centering}
		\includegraphics[width=1.0\columnwidth]{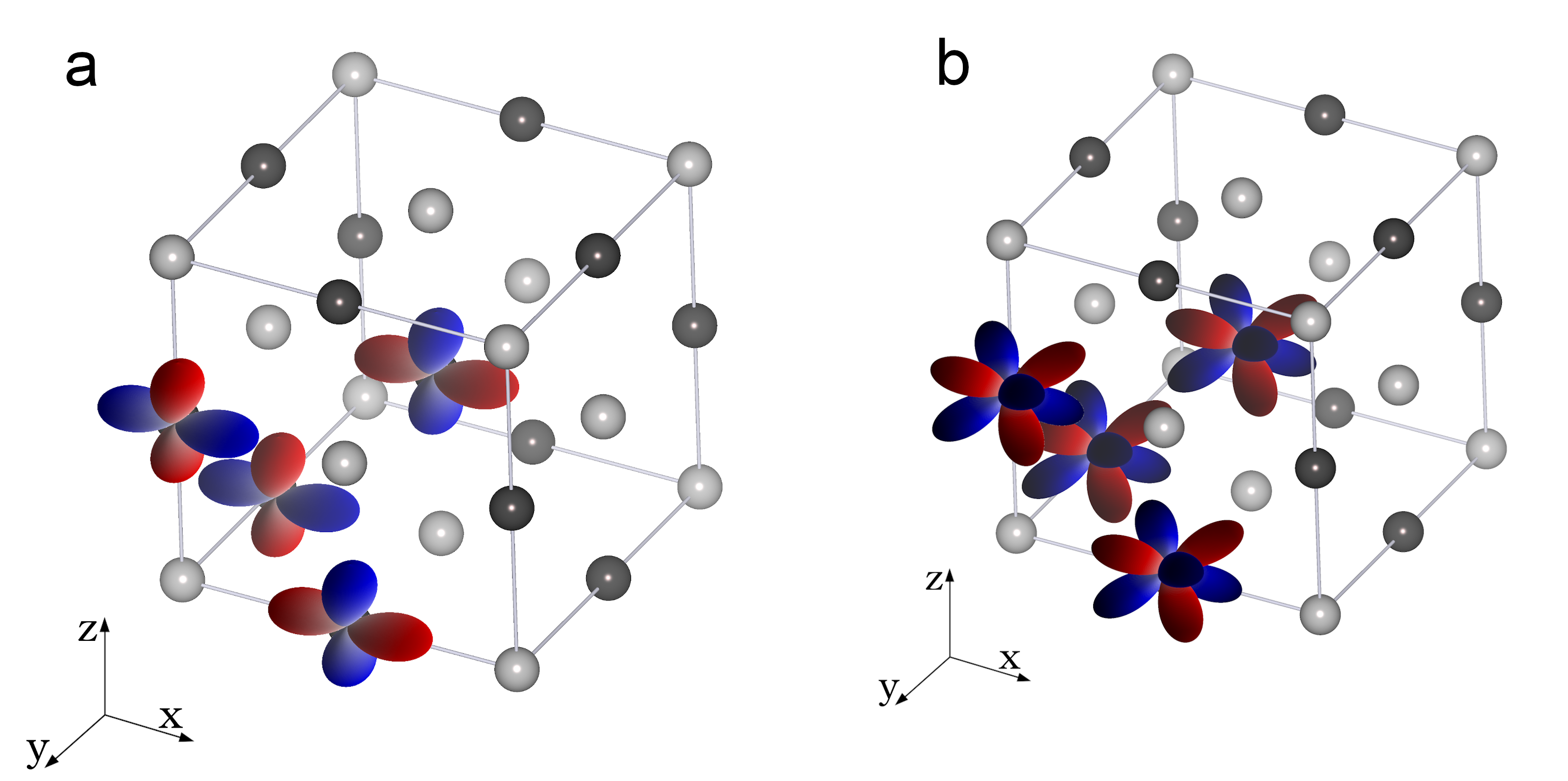} 
		\par\end{centering}
		\caption{(a). Calculated anti-ferro quadrupolar order. Only the Os and $M$ sites (Ca, Mg, or Zn) are shown as dark and light grey balls, respectively. The pattern of ordered quadrupoles is shown at four Os sites forming a primitive fcc unit cell; the quadrupoles at other Os sites are obtained by conventional lattice translations. The quarupoles are shown by polar plots, where the distance from the origin and the color indicates the quadrupole absolute magnitude and its sign. For simplicity, in the present  plot we neglected the hexadecapole contributions to $\tau_{x(z)}$. (b). Calculated ferro-octupolar ground-state order. Polar plots of $xyz$ octupoles are shown for the same Os sites as in panel (a).}	\label{fig:AFQ_order} 
\end{figure}

In Supp. Fig.~\ref{fig:AFQ_order}a we display the calculated planar anti-ferro quadrupole order (AFQ) that is stabilized with the $xyz$ IEI set to zero. This structure is formed by a ferro-quadrupolar order within (001) planes ($yz$ in the Supp. Fig.~\ref{fig:AFQ_order}); those planes are AF-stacked in the perpendicular direction [ ($x$ in the plot). The saturated order parameters, $|\langle \tau_x\rangle|$ and $|\langle \tau_z\rangle|$, are equal to 0.46 and 0.27, respectively, with  sign alternating between the adjacent (001) planes.

An AFQ order for $d^2$ cubic DP  has been previously obtained by Khaliullin {\it et al.}~\cite{khaliullin2021}. The actual model solved in Ref.~\cite{khaliullin2021} is a classical Heisenberg with the quadrupole degrees of freedom mimicked by unit vectors in the $xz$ plane. The calculated ordered structure depicted in their Fig.~4 agrees qualitatively with ours.  Their ordered moments are inclined  within the $xz$ plane indicating non-zero values for both $\langle \tau_x\rangle$ and $\langle\tau_z\rangle$, in agreement with our result.

The  ferro-octupolar ground state predicted by our calculations is depicted in Supp. Fig.~\ref{fig:AFQ_order}b for comparison. 

